\documentclass[preprint2]{aastex6}

\usepackage{graphics,color}
\graphicspath{{}{./}}

\usepackage{natbib}
\newcommand{\lyanv}{Ly$\alpha$+\ion{N}{5}}
\newcommand{\kms}{{\rm km}~{\rm s}^{-1}}
\newcommand{\autompsf}{i_{\rm AUTO} - i_{\rm PSF}}
\newcommand{\pqso}{{\rm\tt PQSO}}
\newcommand{\pqsomidz}{{\rm\tt PQSOMIDZ}}
\newcommand{\pqsolowz}{{\rm\tt PQSOLOWZ}}
\newcommand{\pqsohiz}{{\rm\tt PQSOHIZ}}

\newcommand{\EW}{\mathrm{EW}_0}

\begin{document}

\shorttitle{The $z\sim5$ QLF}

\shortauthors{McGreer et al.}
\title{The Faint End of the $z=5$ Quasar Luminosity Function from the CFHTLS}
\author{Ian D. McGreer and Xiaohui Fan}
\affil{Steward Observatory, The University of Arizona, 
                 933 North Cherry Avenue, Tucson, AZ 85721--0065, USA}
\and
\author{Linhua Jiang}
\affil{The Kavli Institute for Astronomy and Astrophysics, Peking University, Beijing 100871, China}
\and
\author{Zheng Cai}
\affil{UCO/Lick Observatory, University of California, 1156 High Street, Santa Cruz, CA 95064, USA}

\email{imcgreer@as.arizona.edu}

\begin{abstract}
We presents results from a spectroscopic survey of $z\sim5$ quasars in 
the CFHT Legacy Survey (CFHTLS). Using both optical color selection and a 
likelihood method we select 97 candidates over an area of 105 deg$^2$ to
a limit of $i_{\rm AB} < 23.2$, and 7 candidates in the range
$23.2 < i_{\rm AB} < 23.7$ over an area of 18.5~deg$^2$. 
Spectroscopic observations for 43 candidates were obtained with Gemini, 
MMT, and LBT, of which 37   are $z>4$ quasars. This sample extends measurements 
of the quasar luminosity function $\sim$1.5 mag fainter than our previous work 
in SDSS Stripe 82. The resulting luminosity function is in good agreement with 
our previous results, and suggests that the faint end slope is not steep. We 
perform a detailed examination of our survey completeness, particularly the 
impact of the Ly$\alpha$ emission assumed in our quasar spectral models, and 
find hints that the observed Ly$\alpha$ emission from faint $z\sim5$ quasars 
is weaker than for $z\sim3$ quasars at a similar luminosity. Our results 
strongly disfavor a significant contribution of faint quasars to the 
hydrogen-ionizing background at $z=5$.
\end{abstract}


\section{Introduction}\label{sec:intro}

Surveys of high-redshift quasars provide essential insight into the 
growth of supermassive black holes at high redshift and the buildup of 
the metagalactic ionizing background. The Sloan Digital Sky Survey (SDSS)
provided the first large sample of $z\ga5$ quasars \citep{Fan+99,Fan+00}
and measurements of the luminosity function at $z\sim5\mbox{--}6$ 
\citep{Fan+01hiz,Fan+01LF,Richards+06}. The SDSS main survey was successful 
at finding the brightest quasars at high redshift \citep{Jiang+16}, while
the deeper imaging from the multiply scanned Stripe 82 region yielded
fainter quasars \citep{Jiang+08,Jiang+09,McGreer+13}. The sample
of quasars at $z\ga6$ now exceeds one hundred due to additional quasar 
searches such as the Canada-France High-z Quasar Survey \citep{Willott+10}, 
the Subaru High-z Exploration of Low-Luminosity Quasars \citep{Matsuoka+16}, 
and the Pan-STARRS1 Distant Quasar Survey \citep{Banados+16}.

A perhaps surprising result from these surveys is that the most luminous 
quasars --- powered by the most massive black holes --- in general formed 
earliest, with some extreme cases \citep{Mortlock+11,Wu+15} posing a 
challenge to models of black hole formation and growth at early times 
\citep[see reviews by ][]{VB12,Haiman13}.
These luminous systems may have formed in unusual conditions 
\citep[e.g.,][]{VR06,Dijkstra+08,Agarwal+14,BV16}.
On the other hand, faint quasars are better suited to probe lower-mass 
systems at an epoch closer to their original seed mass 
\citep{VLN08,Devecchi+12,Haiman13,Natarajan14,RC16}; 
furthermore, faint quasars capture the rapid buildup of the population of 
massive black holes from $z\sim7$ to $z\sim3$ that likely generates the 
hard ionizing photons required for the reionization 
of \ion{He}{2} \citep[e.g.,][]{McQuinn+09,LT15,DAloisio+17}.
Due to the difficulty of identifying large numbers of faint quasars at
high redshift, the faint end of the luminosity function remains poorly
constrained.

Spectroscopy of high redshift quasars provides highly useful constraints
on the ionization state of intergalactic hydrogen at $z\sim6$, indicating
that reionization has completed by this epoch 
\citep{Fan+06,Becker+14,MMF11,MMD15}. Luminous quasars are too 
rare to provide sufficient photons to be the primary driver of hydrogen 
reionization \citep[e.g.,][]{Fan+01hiz}, but the role played by low-luminosity
AGN remains unknown. Recently, \citet{Giallongo+15} suggested that the
faint number counts are much higher than would be expected from extrapolation
of observations of brighter quasars \citep[but see][]{PDM17}, leading
to renewed interest in AGN-driven reionization models 
\citep[e.g.,][]{MH15}.

Wide area optical surveys are the most fruitful locations to search
for high redshift quasars. First, certain combinations of optical 
colors provide relatively efficient cuts to separate quasars and stars. 
Second, when compared to surveys at other wavelengths (e.g., mid-IR and
X-ray), optical surveys have a more optimal combination of depth and area,
which is required as even at faint fluxes the spatial density of high-$z$
quasars is quite low.

In a previous work we utilized the Stripe 82 region of the SDSS to 
measure the quasar luminosity function at $z\sim5$ to a limit of 
$i_{\rm AB}=22$ \citep[][hereafter Paper I]{McGreer+13}. Stripe 82 is 
a 250~deg$^2$ patch of sky with a coadded image depth $\ga2$~mag fainter 
than the main SDSS survey \citep{Jiang+14}. Using simple color selection 
we produced a sample of 92 $z\sim5$ quasar candidates and obtained 
spectroscopic observations for 73 objects, including 71 quasars at $z>4$. 
From a well defined sample of 52 quasars at $4.7<z<5.1$ we measured the 
luminosity function to a limit of $M_{1450}\approx-24$. Combining this 
sample with bright quasars from the full SDSS footprint, we found that 
the break in the luminosity function occurs at a relatively high 
luminosity ($M_{1450}^* < -27;~2\sigma$), and that the number counts 
increase steeply below this value, with a faint end slope 
$\alpha \approx -2$ 
\citep[compared to $\sim-1.3$ at lower redshift, e.g.,][]{Ross+13}.

The success of the color selection applied to Stripe 82 was due to
the relative ease with which quasars can be distinguished from stars
at $z\sim5$ using optical colors, and the high quality of the Stripe 82 
photometry at $i\approx22$. The CFHTLS Wide provides images in SDSS-like 
$ugriz$ bandpasses to a depth $\sim2$ mag fainter than the Stripe 82 
coadds\footnote{\url{http://www.cfht.hawaii.edu/Science/CFHTLS/}}. 
We thus decided to employ similar methods to the CFHTLS data in order to 
extend our measurements of the faint end of the quasar 
luminosity function at $z\sim5$. In previous works we presented
interesting objects discovered in the course of this survey; namely,
a $z=5$ binary quasar with a projected separation of 21\arcsec\ 
\citep{McGreer+16}, and an unusually bright Lyman-alpha emitter
at $z=5.4$ \citep{McGreer+17}.

All magnitudes are reported on the AB system 
\citep{OG83} and corrected for
Galactic extinction \citep{SFD98} unless otherwise noted.
We use a $\Lambda$CDM cosmology with parameters
$\Omega_\Lambda=0.728$,~$\Omega_m=0.272$,~
$\Omega_{\rm b}=0.0456$,~and~
$H_0=70~{\rm km}~{\rm s}^{-1}~{\rm Mpc}^{-1}$
\citep{Hinshaw+13}, which is updated from the cosmology used in
Paper I \citep{Komatsu+09} although this results in only minor
differences.

\begin{figure*}[!th]
 \epsscale{1.1}
 \plotone{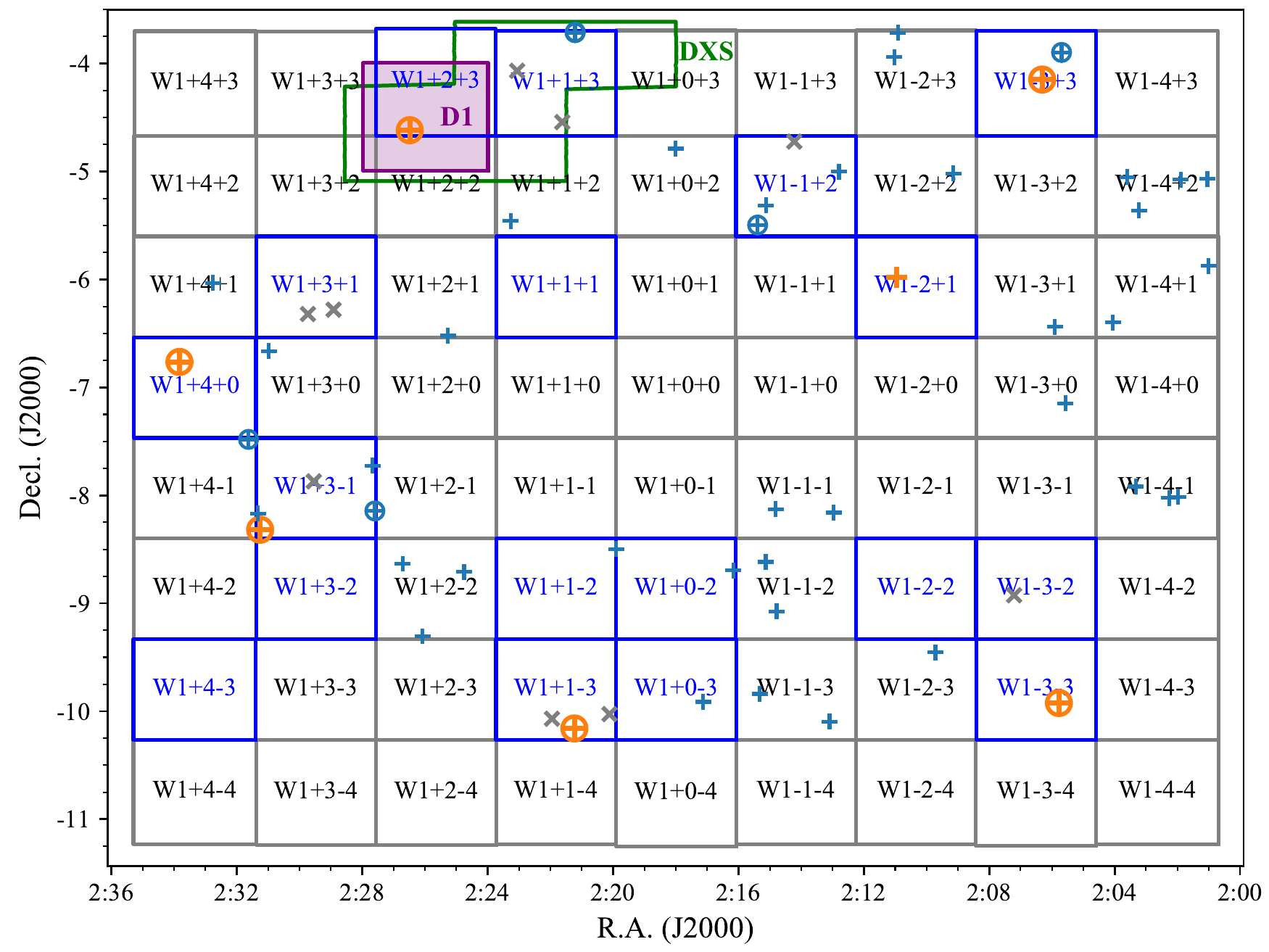}
 \caption{Layout of the CFHTLS W1 field. Individual Megacam pointings are
 labeled as in \citet{Gwyn12}. Gray (blue) lines mark the boundaries of 
 pointings rejected (accepted) by our field selection cuts for the faint 
 candidate selection. The boundary of the UKIDSS DXS region is displayed
 with a green line, and the Deep D1 field as a filled purple square.
 The positions of our final quasar candidates are marked with plus signs,
 with blue (orange) indicating bright (faint) objects and circles drawn around
 objects with spectroscopy. Additional objects with spectroscopic observations 
 not included in the final candidate list are marked with gray crosses.
 \label{fig:cfhtlsw1field}
 }
\end{figure*}

\section{CFHTLS imaging data}

The CFHTLS Wide encompasses a total of 150~deg$^2$ 
split over four widely spaced fields. Stacked images and derived catalogs
from the MegaPipe processing are publicly available on the CADC 
website\footnote{\url{http://www4.cadc-ccda.hia-iha.nrc-cnrc.gc.ca/en/megapipe/cfhtls/index.html}} 
and are described in \citet{Gwyn12}. Not only are the CFHTLS Wide data 
considerably deeper than the Stripe 82 coadded imaging, they also have 
superior image quality. The stacked images have typical 5$\sigma$ magnitude 
limits of $r=25.9$, $i=25.7$, and $z=24.6$, and PSF FWHM $\sim$0.6-1.0\arcsec; 
compared to $i \approx 23.5$ and $\sim$1.0\arcsec\ for Stripe 82.

We also utilize the CFHTLS Deep survey data. These consist of four
independent, single MegaCam pointings (1~deg$^2$ in area) with depths
of $i=27.4$. Two versions of the Deep image stacks are available: the 
``full-depth'' coadds which utilize a larger number of input images, and 
the ``best-seeing'' coadds which are derived from a subset of images with 
the best image quality.

Image stacks for both the Wide and Deep data were generated by selecting 
input images based on quality criteria, resampling the images to remove 
geometric distortions, and combining them with a median algorithm using 
\texttt{SWarp}. The astrometric and photometric calibrations were obtained 
from comparison to SDSS measurements when available. The final catalogs 
were produced with \texttt{SExtractor}. Complete details of the MegaPipe 
processing are provided in \citet{Gwyn12}. The layout of the W1 field
can be viewed in Figure~\ref{fig:cfhtlsw1field}.

As demonstrated in Paper I, color selection of quasars at $z\sim5$ is 
efficient owing to the separation between typical quasar colors at this 
redshift and the track defined by the stellar locus. However, this 
efficiency is extremely sensitive to the photometric accuracy, as the 
population of M dwarf stars similarly with red colors is several 
orders of magnitude more numerous than the high redshift quasar population. 
Furthermore, at increasingly faint fluxes unresolved galaxies with red 
colors begin to outnumber stars, particularly at high Galactic latitudes. 
For these reasons obtaining both an accurate photometric calibration and 
reliable star/galaxy separation from the CFHTLS imaging is crucial to this 
work.

We chose to generate our own photometric catalogs from the CFHTLS Wide
images in order to better understand any systematic variations in the
photometry and to improve the star/galaxy separation. These catalogs
include PSF-based measurements using the PSF kernels obtained from 
\texttt{PSFEx} as part of the MegaPipe processing. Detection is performed on
the $i$-band images and forced photometry is obtained in the other bands.
We perform several diagnostic tests in order to assess the reliability of 
our photometric catalogs and to control for systematics in the CFHTLS imaging, 
as detailed in the following subsections.

\subsection{Star/galaxy separation}\label{sec:stargal}

An initial examination showed that the CLASS{\_}STAR parameter provided in 
the CFHTLS \texttt{SExtractor} MegaPipe catalogs was not reliable in our 
magnitude range of interest ($i\ga23$). After exploring several approaches 
to this problem, we adopt a star/galaxy criterion similar to that employed 
by the SDSS. Namely, we use the difference between the flux measured 
with a PSF-shaped aperture and that measured through an elliptical 
Kron-like aperture (PSF\_MAG and MAG\_AUTO in \texttt{SExtractor}, respectively).
This differs slightly from the SDSS definition, which uses the best-fit
model magnitude instead of an elliptical aperture (where the model is
selected from either a de Vaucouleurs or exponential disk profile, 
\citealt{Stoughton+02}), but we obtain acceptable results with this
definition.\footnote{We experimented with the SPHEROID and DISK model
photometry implemented in \texttt{SExtractor}, but found the results to be much
noiser than using MAG\_AUTO.}

We examine the reliability of our star/galaxy separation with several 
test sets, paying particular attention to the performance at faint 
magnitudes. First, we use HST data from the CANDELS survey 
\citep{Grogin+11,Koekemoer+11}. The UDS ACS/F814W imaging has an area of 
$\sim150$ arcmin$^2$ and lies within the CFHTLS-W1 field. We select 292 
stars from the UDS v1.0 catalogs \citep{Galametz+13} using F814W FWHM
measurements. After matching the stellar objects to the CFHTLS Wide 
catalogs we find that the average $i$-band magnitude difference for stars 
with $i<24$ is $\langle\autompsf\rangle = 0.11$, with 89\% (94\%) of the 
UDS objects having $\autompsf > 0.0~(-0.1)$. Thus this method is highly 
complete at selecting stellar objects from the HST imaging; however, 
these results are based on a small region lying entirely within a single 
Wide pointing (W1+0+2) that has exceptionally good seeing (0.57\arcsec\ 
in $i$-band) and may not be representative of the full survey.

\begin{figure}
 \epsscale{1.1}
 \plotone{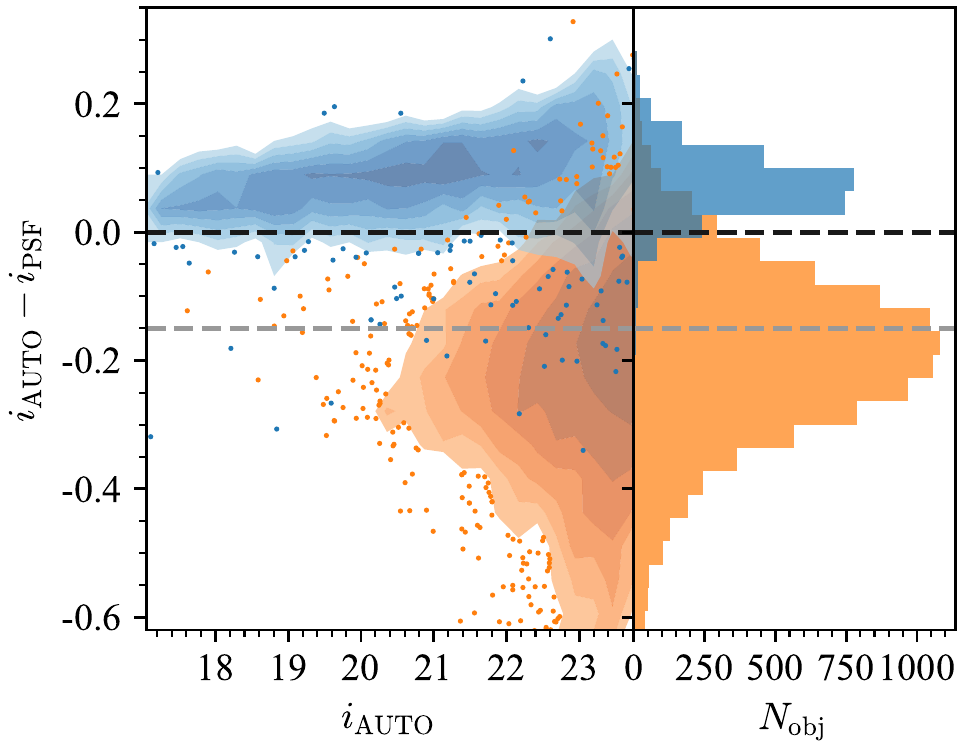}
 \caption{Demonstration of the star/galaxy separation cuts. The Deep D1
 best-seeing catalogs are used as a reference. Stellar objects are selected
 from D1 using a cut on the half-light radius of $r_{1/2}<0.4$\arcsec. Compact
 galaxies are selected in the range $0.45$\arcsec$<r_{1/2}<0.55$\arcsec. Both 
 samples are cross-matched to counterparts in the overlapping Wide catalogs. 
 In the left panel the blue (orange) shaded contours and points represent the 
 density of stellar (non-stellar) objects as a function of magnitude and 
 MAG\_AUTO$-$MAG\_PSF. The right panel displays histograms of both samples.
 The black dashed line is the more strict cut used for the faint candidate
 selection, which is $\sim88$\% complete at $i_{\rm AB}\ga23$ while rejecting
 $>95$\% of the compact galaxies. The more inclusive cut used for the bright
 candidate selection is displayed with a gray dashed line and is $\sim98$\%
 complete to faint stellar objects.
 \label{fig:stargal}
 }
\end{figure}

For a second test we use the Deep stacks from the CFHTLS. The Deep field 
D1 lies entirely within the Wide field W1 and overlaps with four 
individual Wide pointings. The D1 best-seeing stacks have an image quality 
of 0.64\arcsec\ in the $i$-band, compared to 0.7-0.9\arcsec\ in the
overlapping Wide fields. We utilize the superior depth and resolution of
the Deep data to identify stellar objects and then compare to the  
corresponding Wide photometry.

We use the MegaPipe D1 best-seeing catalogs to select stars from the 
locus of points in size-magnitude space. We find that the stellar locus 
is resolved from the galaxy distribution to a limit of $g \la 23.7$. We 
use the $g$-band to select stellar objects, then examine their $i$-band 
photometry for classification.
When the cut $\autompsf > 0.0~(-0.15)$ is applied to the Wide catalogs, 
94\% (99\%) of the D1 stars are selected to a limit of $i=23.7$, in good 
agreement with the HST results. Focusing on the range $22.8<i<23.7$, the 
completeness is slightly lower, with the same cuts selecting 88\% (98\%) 
of the unresolved sources from D1. If we select compact galaxies by 
identifying marginally resolved sources with a measured size just above 
the stellar locus ($r_{1/2} \approx 0.5$\arcsec, compared to 
$r_{1/2} \approx 0.35$\arcsec\ for stars), a cut of $\autompsf > 0.0$ 
eliminates $\sim96$\% of the compact galaxy contamination at  $i \ga 23$. 
The results of this test are presented in Figure~\ref{fig:stargal}.

Finally, we check these cuts against a sample of known quasars drawn 
from our spectroscopy of candidates in the CFHTLS. In early versions of 
our candidate selection weaker star/galaxy cuts were applied, 
resulting in greater contamination. Half of the non-quasars from our
spectroscopic observations are rejected by an $\autompsf>0$ cut.

We conclude from these tests that a cut of $\autompsf > 0.0$ is 
$\sim90\%$ complete at selecting stellar objects from the Wide imaging,
while greatly reducing the contamination from compact galaxies. We adopt
this cut for selecting faint quasars, where the galaxy contamination is
greater, while using the more permissive $\autompsf > -0.15$ cut for
brighter objects.

\begin{figure}
 \epsscale{1.1}
 \plotone{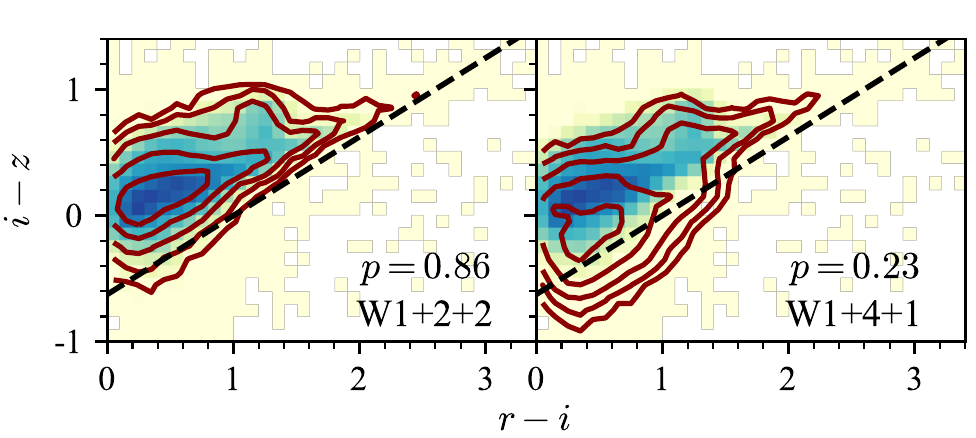}
 \caption{Examples of the stellar locus matching procedure used to select
 fields with greater photometric reliability. In both panels the background
 image represents the density of stars in $r-i$/$i-z$ color space in the Deep 
 D1 field, with blue (yellow) indicating higher (lower) density. The dark red 
 contours denote the stellar density in two different Wide pointings. The left 
 panel presents a pointing that is an example of a good match to the Deep 
 photometry, while the pointing in the right panel is a poor match.
 For reference, the black dashed line marks the color cut used to separate 
 stars and quasars in Paper I; the pointing on the right has substantially
 greater contamination from stars scattered below this boundary.
 \label{fig:systematics}
 }
\end{figure}

\subsection{Field selection}\label{sec:fieldselection}

The expected density of $z\sim5$ quasars with $i\la23$ on the sky is
$\sim1~{\rm deg}^{-2}$. Searching the full CFHTLS Wide area (150~deg$^2$)
would result in too many candidates to confirm spectroscopically given
30-60 min exposures for the faintest targets. We thus focus our attention 
on individual pointings within the CFHTLS that are likely to yield the 
highest reliability for color selection. In this section we detail the 
criteria used to select CFHTLS pointings in order to define our survey 
area. As described in \citet{Gwyn12}, each pointing is a contiguous, 
$\sim1~{\rm deg}^2$ area corresponding to a single MegaCam field-of-view, 
with a small overlap area between adjacent pointings.

First, we require homogeneous $i$-band filter coverage. During the course 
of the CFHTLS the $i$-band filter was replaced. The two filters used, 
$i^\prime_1$ and $i^\prime_2$, have significantly different profiles at 
the blue edge of the filter. As can be seen in Fig.~6 of \citet{Gwyn12}, the 
$i^\prime_2$ filter has a peak $\sim150$\AA\ bluer than the $i^\prime_1$ 
filter. This shift has a substantial effect on quasar colors at $z\sim5$, 
inducing differences of $\sim0.1$--$0.2$~mag in the $r-i$ and $i-z$ 
colors between the two filters. As most of the CFHTLS was observed with the 
$i^\prime_1$ filter, we simply remove all fields with $i^\prime_2$ coverage 
from our survey in order to maintain a consistent set of selection criteria.

Next we select fields in order to optimize the photometric reliability.
The five-band imaging for the CFHTLS was performed over a period of years,
with varying conditions occurring in the individual images contributing
to the final stacked images. This can lead to difficulties in obtaining
an accurate representation of the coadded PSF, as well as non-uniform
depths between the different bands.

To address the issue of non-uniform depth, we utilize the limiting depths
for each pointing and each band given in Table 4 of \citet{Gwyn12}. For our
faint object selection criteria (defined in \S\ref{sec:widetarg}), we require 
that the limiting magnitudes (50\% completeness) in the bluest and reddest 
bands are $g_{\rm lim}>26.3$ and $z_{\rm lim}>24.5$, respectively. These two 
bands are crucial in $z\sim5$ quasar selection. The Lyman Limit for $z\sim5$ 
quasars is redward of the $g$-band, thus relatively deep $g$-band imaging 
ensures that candidates are $g$-band dropouts. The $z$-band is typically 
much shallower than the $i$-band, but blue $i-z$ colors are an important
discriminant between quasars and stars, and thus reliable $z$-band photometry 
is a necessity.

To further constrain the photometric reliability we assess the colors of 
objects along the stellar locus. Our approach is broadly similar to 
using principle colors to define a well calibrated stellar locus
\citep{Ivezic+04} or stellar locus regression \citep{High+09}. However, 
as we are only interested in selecting the best fields we do not attempt to 
improve the calibration \citep[cf.][who discuss a recalibration of the CFHTLS
data]{Matthews+13}. We define the stellar locus using photometry from
the Deep survey as a reference, and then compare photometry from individual
Wide pointings to this reference using the binned maximum likelihood (ML) 
as a metric, where the data are binned in their $r-i$ and $i-z$ colors. 
Figure~\ref{fig:systematics} presents examples of this method. The left 
panel contains a color-color plot from a ``good'' Wide pointing with colors 
well matched to the Deep catalogs. A poor match is presented in the right 
panel. The binned ML provides the probability of a match between the two 
distributions; based on examining the color-color plots we select a threshold 
of $p_{\rm locus}>0.52$, which removes $\sim40$\% of the W1 pointings. 
Imposing all of the quality cuts on W1 retains only 18/72 (25\%) of 
the pointings in the full area; we will utilize this region for selection of 
the faintest candidates (\S\ref{sec:widetarg}).

\section{Target selection}

\subsection{Simulated quasar colors}\label{sec:simqso}

As in Paper I, our target selection is guided by models for quasar colors at
$z\sim5$. The paucity of known quasars at this redshift means that a limited
training set is available, we thus generate simulated quasar colors trained
on the properties of the more abundant quasar population at lower redshift,
assuming that quasar spectra do not strongly evolve \citep[e.g.,][]{Kuhn+01,Jiang+06}.
Specificially, our model is derived from the observed spectral properties of
SDSS BOSS DR9 quasars \citep{Ahn+12,Ross+13}, which capture the UV emission of
$\sim150,000$ quasars at $z\sim2.5$ over a wide luminosity range. The model
includes a broken power law continuum, an emission line template capturing the
Baldwin Effect \citep{Baldwin77}, iron emission templates 
\citep{BG92,VW01,Tsuzuki+06}, and a stochastic IGM HI absorption model 
\citep{WP11,McGreer+13}. Quasar spectra are generated though Monte Carlo 
samplings of the individual spectral features in order to reproduce the 
intrinsic scatter in colors at any given redshift. The simulated spectra are 
then convolved with filter bandpasses to produce mock photometry, and then 
realistic scatter is introduced to mimic actual observations. The code to 
generate simulated quasar spectra and photometry is written in Python and is 
freely available\footnote{\url{https://github.com/imcgreer/simqso}};
additional details of its implementation are provided in Paper I and
\citet{Ross+13}.

\begin{figure}
 \epsscale{1.1}
 \plotone{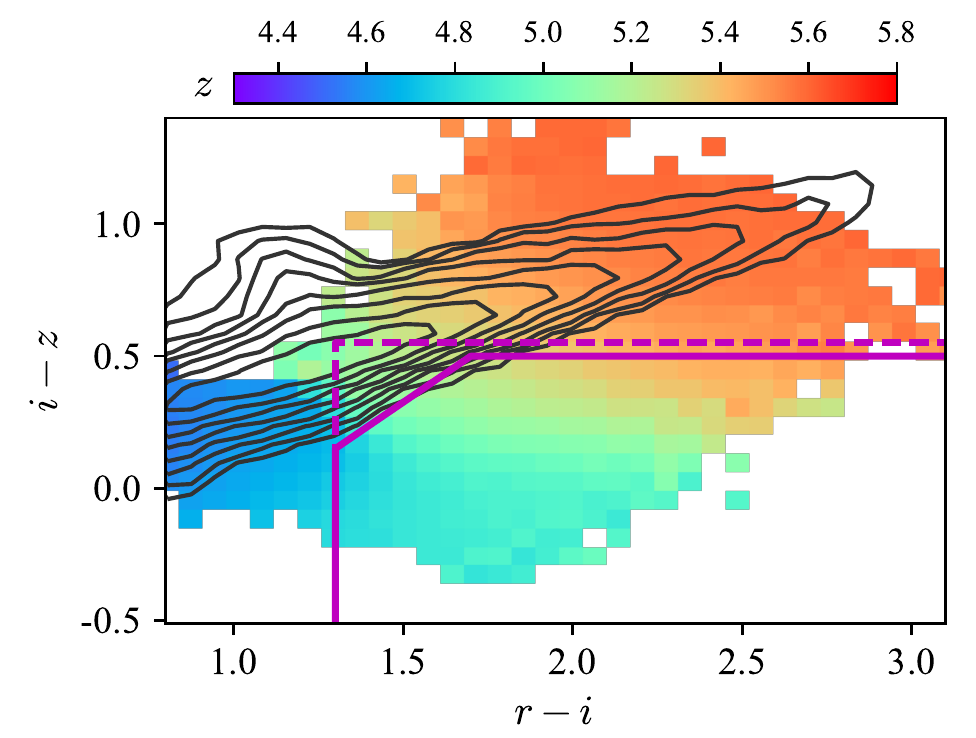}
 \caption{Simulated quasar colors in $riz$ color space. The simulations
 include 140,000 objects in the range $4.3<z<5.8$. The color scale indicates
 the mean redshift in bins of width $0.07$~mag in $riz$ colors. The location 
 of the stellar locus is given by logarithmically stepped contours represented 
 by dark gray lines. The color selection efficiency peaks at $z\sim5.0$ as the
 colors of typical quasars separate from the stellar locus.
 \label{fig:simcolors}
 }
\end{figure}

We adopt the model parameters used in Paper I, but update the model to
include photometry in the CFHT MegaCam system with photometric
errors appropriate for the CFHTLS Wide survey. Results of the simulations are
presented in Figure~\ref{fig:simcolors}, demonstrating that for our fiducial
quasar model the $riz$ colors of $z\sim5$ quasars are well separated from the 
stellar locus.

\subsection{Color selection}\label{sec:colorsel}

The color criteria presented in Paper I were designed for quasar selection 
in Stripe 82 and require modification for the present work: although the 
CFHT photometric system is quite similar to the SDSS, the differences are 
sufficient to result in markedly different colors of $z\sim5$ quasars (of 
order $\sim0.2$ mag). We adopt the following color cuts for the CFHTLS 
Wide selection:

\begin{equation}\label{eqn:strict_usnr}
 S/N(u) < 2.2 ~~,
\end{equation}
\begin{equation}\label{eqn:strict_gsnr}
 g-r > 1.8 ~~{\rm OR}~~ S/N(g)<2.2 ~~,
\end{equation}
\begin{equation}
 r-i > 1.3 ~~,
\end{equation}
\begin{equation}\label{eqn:strict_izcut1}
 i-z < 0.15 + 0.875[(r-i) - 1.3] ~~,
\end{equation}
\begin{equation}\label{eqn:strict_izcut2}
 i-z < 0.5 ~~.
\end{equation}

We refer to these cuts at the ``strict'' color criteria.

Alternatively, we also consider ``weak'' criteria defined by replacing 
equations ~\ref{eqn:strict_izcut1} ~and~ \ref{eqn:strict_izcut2} with:
\begin{equation}\label{eqn:loose_izcut}
 -0.5 < i-z < 0.55 ~~.
\end{equation}
The weak criteria are highly complete but by themselves result in an 
unacceptably high level of contamination. These criteria will be used when 
ancillary data is available to reduce the contamination.

In Paper I we utilized near-IR imaging from UKIDSS to assist in rejecting
stars. The CFHTLS Wide fields are not uniformly covered by sufficiently deep
near-IR imaging to aid in quasar candidate selection. 
Alternative means of expanding on optical color selection, e.g., through
radio \citep{MHW09} or mid-IR \citep{Wang+16} data were similarly not
available to sufficient depth and area in these fields. We thus found it 
desirable to prioritize the color-selected candidates with a probabilistic 
approach.

\subsection{Likelihood selection}\label{sec:likesel}

We adopt the likelihood method \citep{Kirkpatrick+11} to rank our candidates 
and to provide additional candidates missed by the color cuts.
This method requires a training set of quasars (``QSO'') in the desired
redshift range, and a catalog of non-quasar contaminants (``Everything Else``,
or EE). After properly normalizing the training set catalogs, quasar 
probabilities are assigned to input objects by asking whether they are more 
likely to belong to the QSO or EE catalogs. The relative probabilities are 
obtained from a $\chi^2$ statistic calculated by comparing the input fluxes
and errors to the training set fluxes in a multidimensional space.

The likelihood method has the limitation of including photometric errors in 
the training sets (see \citealt{Bovy+11} for a related method that avoids this 
issue). We mitigated this problem by using the CFHTLS Deep survey catalogs,
which have substantially smaller photometric errors than the Wide catalogs,
for our EE training set and the (noise-free) simulated quasar photometry for
our QSO training set. The simulated quasars were necessary as no training set 
of known quasars at $z\sim5$ with similar fluxes as our targets exists.

\begin{figure}
 \epsscale{1.1}
 \plotone{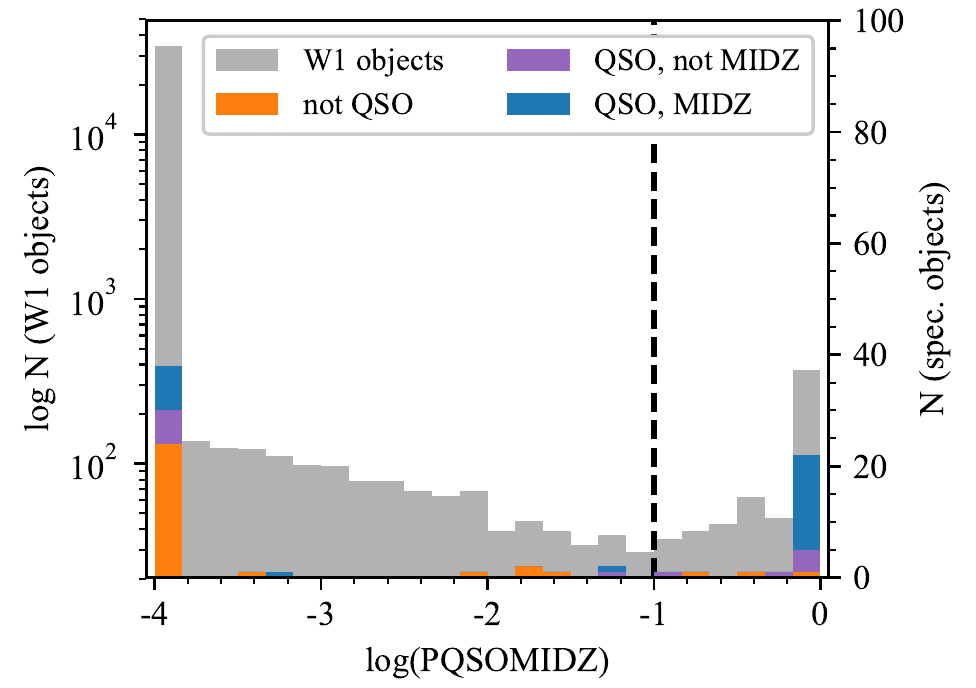}
 \caption{Distribution of log-likelihoods that objects are quasars at
 $4.8<z<5.2$ ($\pqsomidz$). The gray histogram is constructed from the W1
 pre-selected object catalog (\S\ref{sec:likesel}) and is shown on a
 logarithmic scale (left axis). The number of objects with spectroscopy 
 is tallied as orange (non-quasars), purple (quasars not in mid-$z$ range), 
 and blue (quasars in the mid-$z$ range) bars on a linear scale given on 
 the right axis. A lower bound of $10^{-4}$ is set on the probabilities.
 A cut of $\pqsomidz>0.1$ is used to select candidates with the likelihood
 method.
 \label{fig:likelihoods}
 }
\end{figure}

{\bf EE catalog:}
The Everything Else catalog was constructed from the $i$-band detection 
catalogs derived from the full-depth Deep coadds. The Deep field photometry 
is in an identical system as the Wide field, but the photometric uncertainties 
are  negligible in our range of interest.  On the other hand, the Deep fields 
cover a much smaller area and have a limited sampling the full distribution of 
contaminants, particularly at the bright end.

We first restrict the catalog to the unmasked survey regions, resulting in 1.5 
million objects. We then remove objects with suspect photometry (I\_FLAGS=0 
from \texttt{SExtractor}), and apply a cut on stellarity. Formally, our 
likelihood model could include the probability that a candidate object is a 
star, and thus fold in the stellar probability for objects in the EE catalog. 
We instead take the simpler approach of applying the star/galaxy cut 
$\autompsf>-0.15$ to the Deep photometry. We account for inaccuracy in the 
star/galaxy classification by including a random sampling of objects from 
the Deep catalogs that pass the stellarity cut based on their Wide 
measurements but not with the Deep photometry. 
The final EE catalog has 600,000 entries and an effective area of 3.3~deg$^2$.

{\bf QSO catalog:}
The mock quasar catalog is constructed from the simulations. First, we 
distribute a sample of quasars in luminosity and redshift according to the 
luminosity function from Paper I (row 1 in Table 5). The total number matches
the expectation for a survey with an area of 250~deg$^2$, roughly twice the 
area we searched in CFHTLS, and the sample is bounded by $M_{1450} < -21.9$
($i_{\rm AB}<24.2$ at $z=5$) and $4.5 < z < 5.8$.

We generate simulated spectra and photometry as described in 
\S\ref{sec:simqso}. Although the simulated photometry is noiseless, this
approach is limited by any systematic errors in our quasar model. The model
has been shown to accurately reproduce the colors of SDSS BOSS quasars at
$2.2 < z < 3.5$ \citep{Ross+13}, and we assume it applies equally well to
$z\sim5$ quasars. Following \citet{Bovy+11}  we divide the 
simulated quasars into three redshift bins in order to gain information about
objects likely to be just below or above our target redshift range. 
The probabilities are calculated as $p(f_i, z_1 \le z < z_2~{\rm quasar})$ 
where $f_i$ represents the (linear) fluxes in the five SDSS bands and the
three redshift bins are $\pqsolowz$\ ($4.5<z<4.8$), $\pqsomidz$\ ($4.8<z<5.2$), 
and $\pqsohiz$\ ($5.2<z<5.8$). We also calculate the summed probability that 
an object is in the full redshift range ($\pqso$). Quasars outside of this 
redshift range are ignored; they will contribute negligible contamination 
given our color cuts, and are implicitly included in the EE catalog.

\vspace{12pt}

With our training sets in hand we now derive quasar probabilities for 
objects in the Wide catalogs. To ease the calculation we only compute 
likelihoods for objects with I\_FLAGS=0, $i<24$, $g-r>1.5$, and $r-i>1.0$.
These criteria are unlikely to miss any quasars at $z\sim5$ and reduce
the input list to $\sim$100,000 objects in the four CFHTLS-Wide fields.
Figure~\ref{fig:likelihoods} presents the distribution of likelihoods for 
the pre-selected list in the W1 field. The vast majority have 
$\pqsomidz < 10^{-4}$. The choice of $\pqsomidz \ga 0.1$ roughly picks 
out a minimum in the distribution separating the likely quasars from the 
stellar contaminants. We thus apply the cut
\begin{equation}
 \pqsomidz > 0.1 
\end{equation}
to define ``likelihood-selected'' candidates. 

\subsection{CFHTLS Wide targets}\label{sec:widetarg}

The final target selection is obtained from a combination of the color
and likelihood selection methods. We define three distinct samples
of targets. First, we select bright candidates ($i_{\rm AB}<23.2$) in
the W1, W3, and W4 fields\footnote{We also have a candidate list for the
W2 field, but we ignore it here as we obtained a spectrum for only a single
object in this field.}. In the W4 field we only considered pointings that
overlap with the SDSS Stripe 82 imaging in order to prioritize candidates
that are in common with the Paper I sample. The bright targets provide 
backup targets for observing runs with sub-par conditions, and an 
independent sample that spans the luminosity range of the Paper I 
QLF, while also extending 1~mag deeper.

\begin{figure}
 \epsscale{1.1}
 \plotone{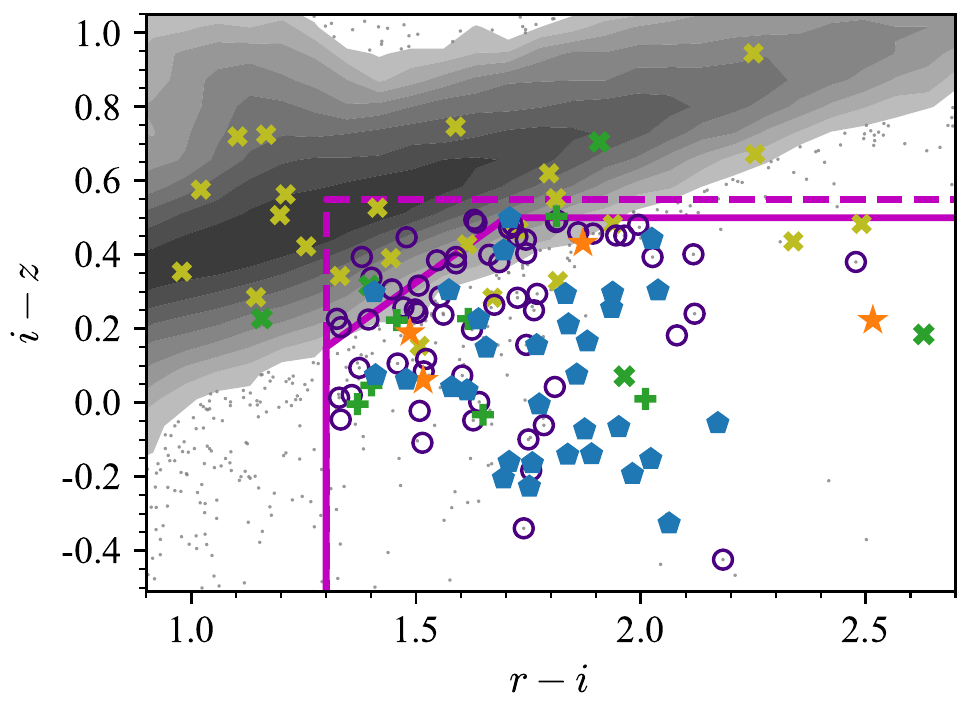}
 \caption{Colors of high redshift quasars and stars in CFHTLS-W1.
 The gray contours and points represent the density in logarithmic steps 
 of catalog objects with $\autompsf > -0.15$ and to a depth of 
 $i_{\rm AB} < 23.2$.
 The solid (dashed) magenta line marks the boundaries of our strict (weak)
 $riz$ color selection criteria.
 Objects in the bright candidate sample are represented by filled blue
 pentagons for spectroscopically confirmed quasars, orange stars for 
 non-quasars, and empty purple circles for objects without spectroscopic
 observations.
 Objects with spectroscopy not included in the final sample are represented
 by green pluses for quasars in the redshift range $4.7<z<5.3$, green
 crosses for quasars outside of that redshift range, and yellow crosses for 
 non-quasars.
 \label{fig:brightcolors}
 }
\end{figure}

For the bright targets we ignore the field selection criteria from 
\S\ref{sec:fieldselection} and select candidates with both the strict 
color criteria from \S\ref{sec:colorsel} and the likelihood selection 
from \S\ref{sec:likesel} (i.e., a candidate may be selected by either 
method). We apply a highly inclusive star/galaxy cut of 
$\autompsf > -0.15$, which is $\ga99$\% complete to stellar objects
(\S\ref{sec:stargal}).  Finally, we visually inspect the candidates and 
discard spurious objects and those with unreliable photometry.

Applying these criteria to the W1, W3, and W4 fields yields 97 
candidates over an area of 105 deg$^2$, after removing 22 during the
visual inspection step. The strict color criteria select 87 of the
candidates, the likelihood criteria 55, and 45 candidates pass both
criteria. Only 10 of the candidates lie outside of the color boxes
and are selected by likelihood only; all of these targets lie within
the weak color criteria. The full list of bright candidates is provided
in Table~\ref{tab:specbright}; a visualization of the color selection of
the bright targets is presented in Figure~\ref{fig:brightcolors}.

\begin{figure}
 \epsscale{1.1}
 \plotone{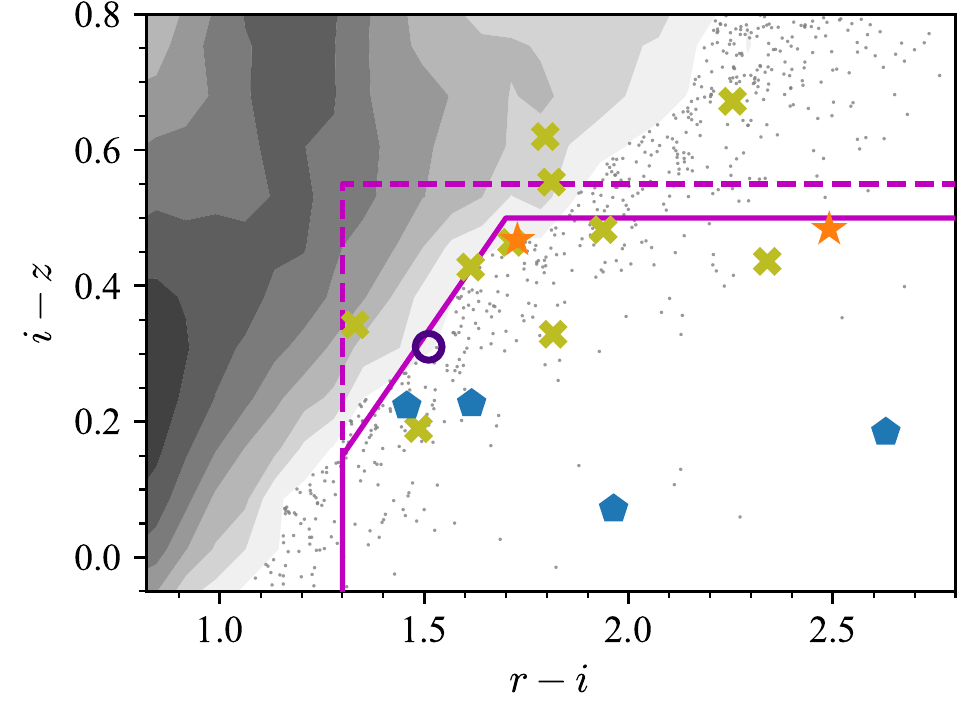}
 \caption{Same as in Fig.~\ref{fig:brightcolors}, but for the faint
 targets in W1. The background contours represent W1 objects with
 $23.2<i_{\rm AB}<23.7$ and $\autompsf > -0.15$; note that the distribution 
 has a somewhat different shape than for the brighter objects likely due 
 to a larger contribution from compact galaxies. A few objects that lie 
 outside the color boxes were observed with Gemini because they were selected 
 as backup targets based on aperture photometry; none of these were found to
 be quasars. Additionally, none of the backup targets selected by colors but
 with low likelihood values were quasars.
 \label{fig:cfhtlsw1riz}
 }
\end{figure}

Next we define the faint quasar sample in the W1 field\footnote{W1 was
selected for its visibility during the Gemini observing run.}, selecting
quasar candidates with $23.2 < i_{\rm AB} < 23.7$. For the faint
targets we apply the field selection criteria to identify the most 
reliable pointings within W1, and we apply a stricter star/galaxy
cut of $\autompsf > 0.0$. This results in seven candidates from an
area of 18.5~deg$^2$, after removing two objects by visual inspection. 
These objects provided the primary target list
for the Gemini observations. The candidate list is provided in
Table~\ref{tab:specfaint}. A color-color plot of the faint candidates 
is presented in Figure~\ref{fig:cfhtlsw1riz}, showing that all seven meet 
the strict color criteria.

\subsection{Ancillary selection in D1+DXS}\label{sec:d1dxs}

Our candidate selection from the CFHTLS-Wide is complemented by deep data
from two ancillary regions. First, the Deep D1 field lies entirely within the 
Wide W1 field, spanning one square degree and overlapping with four separate 
Wide pointings. The D1 $i$-band data are $\approx1.6$ mag deeper on average 
than the overlapping Wide pointings, and the image quality of the best-seeing 
stack is 0.64\arcsec, compared to $<{\rm IQ}(i)>=0.73$\arcsec\ for the Wide 
images.

Second, the W1 field is partially covered by near-infrared imaging from the 
UKIDSS Deep Extragalactic Survey (DXS). As shown in Paper I, near-IR 
photometry is highly useful for rejecting stellar contaminants in $z\sim5$ 
quasar selection, as stars have redder $i-J$ colors than quasars with similar optical colors. At the time of our observations, the DR9 release from UKIDSS 
provided $2.52~{\rm deg}^2$ of $JK$ imaging within the W1 field to
depths of $J_{\rm AB} \approx 23.2$ and $K_{\rm AB} \approx 22.6$ ($5\sigma$).
An area of 0.88~deg$^2$ within DXS overlaps with D1 
(see Fig.~\ref{fig:cfhtlsw1field}).

We first search within the DXS region by applying the weak color criteria from
\S\ref{sec:colorsel} to the W1+DXS overlap area. We remove all cuts on 
morphology, photometric flags, and imaging masks. We then apply the following 
near-IR criteria, similar to those used in Paper I:
\begin{equation}
 i-J < ((r-i)-1.3) + 0.8 ~~,
\end{equation}
\begin{equation}
 i-J < 1.2 ~~,
\end{equation}
after substituting a nominal detection limit for the $J$-band non-detections. 
These cuts select 14 candidates with $23.2<i<23.7$, of which six are within 
D1. Examining the D1 photometry for those six objects, we find that with the 
deeper photometry all but one fail the weak color criteria, and thus are not 
likely to be $z\sim5$ quasars.

We additionally search the 0.88~deg$^2$ D1+DXS overlap region using the D1 
photometry as input to the weak color criteria. This test would identify 
objects with quasar-like colors scattered outside of the color boxes in the 
shallower Wide imaging. This search yields only two candidates, both of which 
were also selected by the W1+DXS search. 

One of the two D1+DXS color-selected candidates has $\autompsf = -0.4$ 
in the D1 best-seeing catalogs. This object is clearly elongated in the 
$i$-band best-seeing image and its radial profile is more extended than 
the profiles of nearby stars. Thus we reject it as a quasar candidate, and 
retain a single good candidate from the D1+DXS area. This candidate is a 
confirmed $z=4.9$ quasar (\S\ref{sec:spectroscopy}).

We also test somewhat more restrictive near-IR color cuts:
\begin{equation}
  i-J < 0.75((r-i)-1.3) + 0.7 ~~,
\end{equation}
\begin{equation}
  i-J < 1.0 ~~.
\end{equation}
This reduces the W1+DXS sample to two objects. One is the quasar also 
identified in the D1+DXS search. The other is also in D1 but falls outside
of the weak color cuts when using the D1 photometry. The fact that nearly
all of the W1+DXS candidates selected with relaxed optical color criteria 
applied to the Wide photometry are rejected either by the more restrictive 
near-IR cuts and/or by using deeper optical photometry suggests that they 
are unlikely to be quasars.

In summary, the D1 and DXS data demonstrate that our selection criteria from
the Wide survey are not missing a significant number of valid candidates 
just outside the selection boundaries. Even after applying highly permissive
color criteria to the Deep catalogs, we identify only a single quasar 
candidate in the D1 area. This corresponds to a sky density of 
$0.40~{\rm deg}^{-2}$, in excellent agreement the result from the Wide 
fields over the same magnitude range, $0.39~{\rm deg}^{-2}$. We kept a 
small number of the W1+DXS-selected candidates in our target list but at
a low priority for observations.

\section{Spectroscopic Observations}\label{sec:spectroscopy}

\begin{figure*}
 \epsscale{1.0}
 \plotone{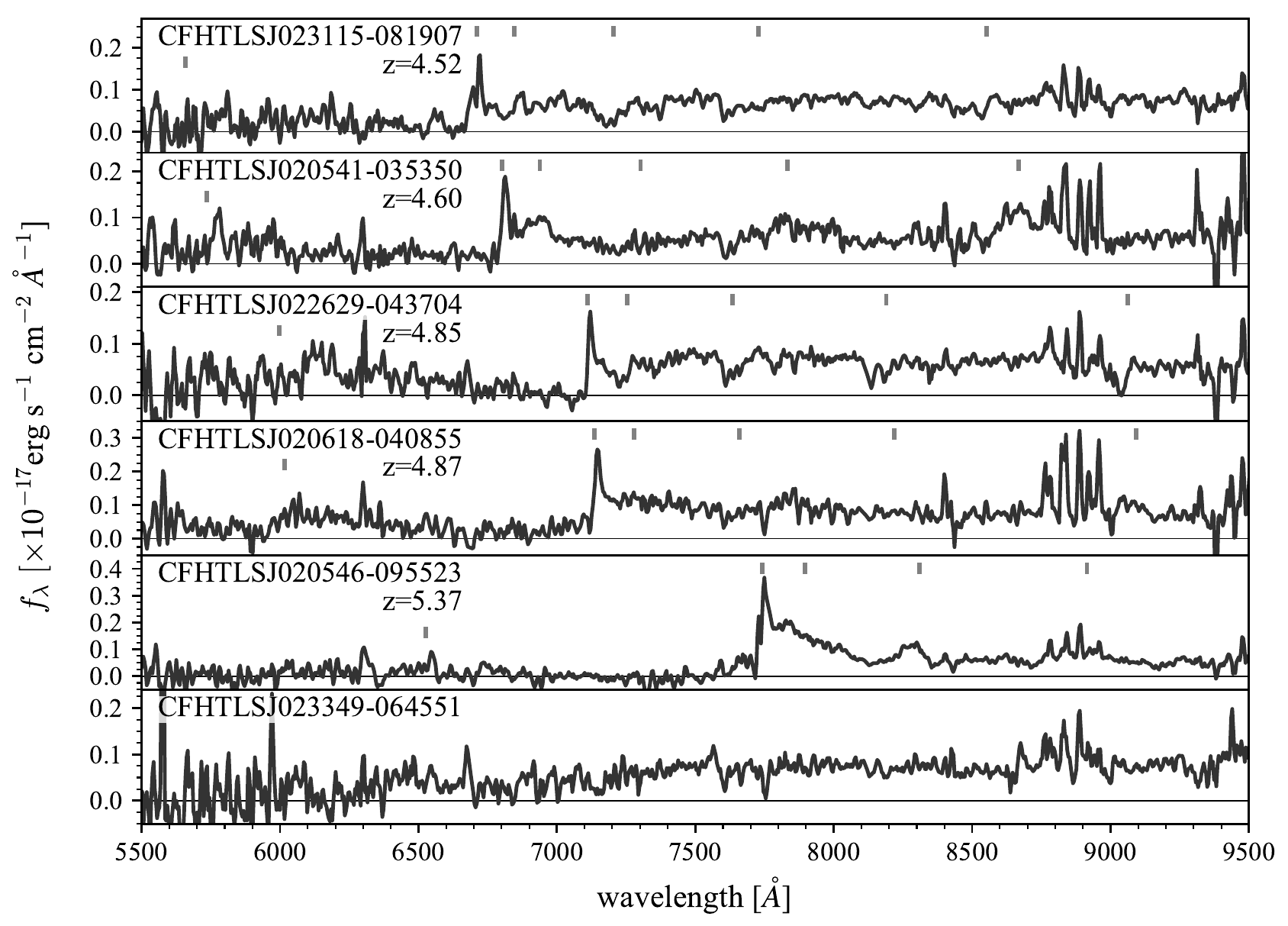}
 \caption{Spectra of CFHTLS quasars observed with GMOS at Gemini North.
 The spectra have been smoothed with a $\sim12$\AA~boxcar. Vertical tick
 marks indicate the location of typical quasar emission lines, from left
 to right Ly$\beta$, Ly$\alpha$, \ion{N}{5}, \ion{O}{1}, \ion{Si}{4}, and
 \ion{C}{4}.  CFHTLSJ020541-035350 is not included in the final candidate
 sample, and CFHTLSJ023349-064551 is an example of a non-quasar.
 The spectra generally display broad emission lines typical of luminous 
 quasars, as well as BAL-like absorption features near the wavelengths of
 \ion{N}{5}, \ion{Si}{4}, and \ion{C}{4}.
 \label{fig:geminispec}
 }
\end{figure*}

\begin{figure*}
 \epsscale{1.1}
 \plotone{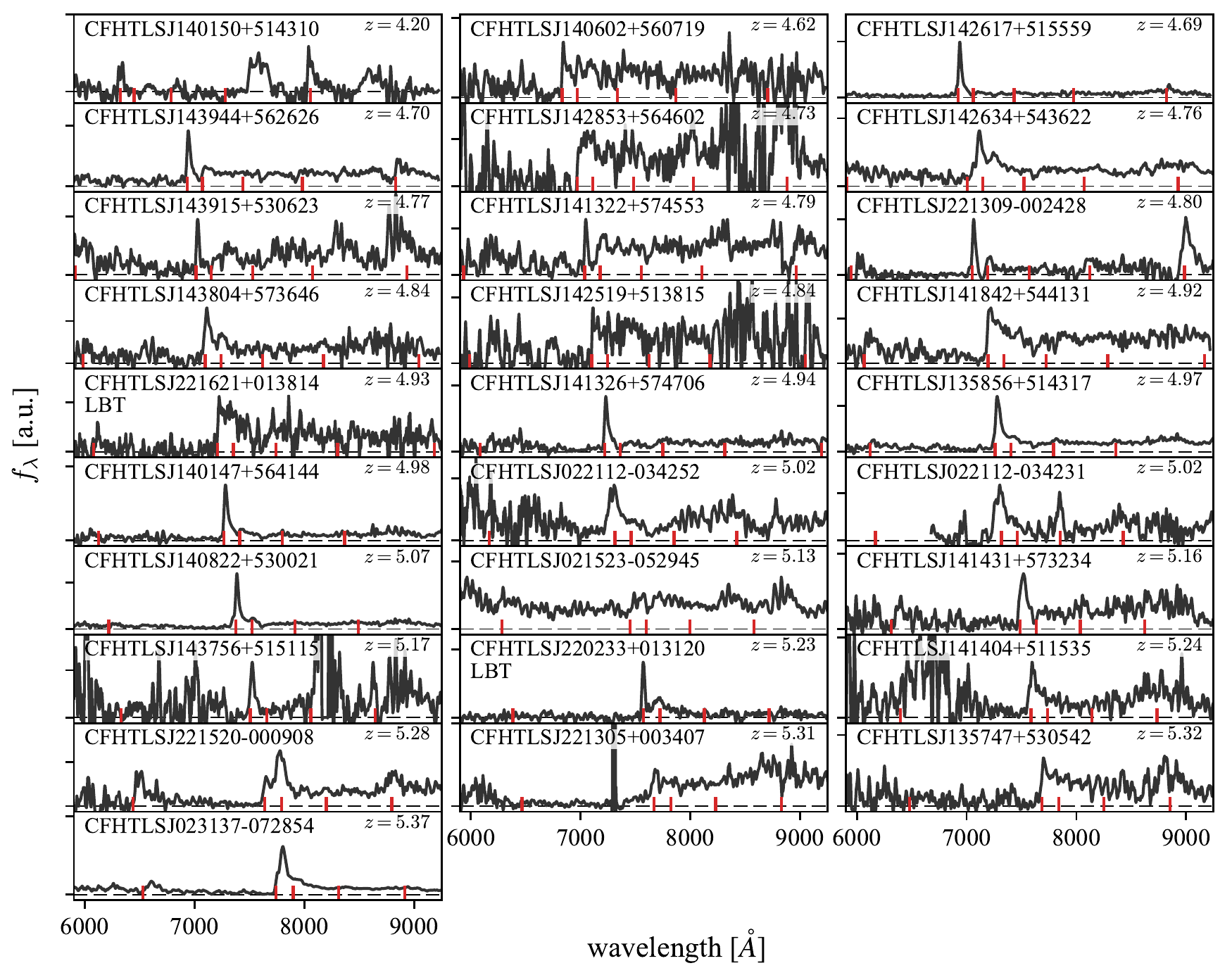}
 \caption{MMT Red Channel spectra of CFHTLS quasars (the two spectra from
 LBT are labeled under the object names). Redshifts are obtained from template
 fitting; however, given the low quality of the spectra and the strong
 absorption features, the redshifts are only
 accurate to $\Delta{z} \sim 0.1$. The same emission lines as in 
 Fig.~\ref{fig:geminispec} are denoted with vertical red marks.
 CFHTLSJ021523-052945 has a noisy extraction but the Ly$\alpha$+\ion{N}{5}
 feature is apparent in the 2D spectrum.
 \label{fig:mmtlbtspec}
 }
\end{figure*}

\subsection{Gemini-North}\label{sec:gemini}

We obtained spectroscopic observations of faint quasar candidates from 
CFHTLS-W1 using GMOS on Gemini North through classical mode observations
conducted on 2013 October 24-25 (program GN-2013B-C-1).
Conditions were excellent with 0{\farcs}4-0{\farcs6} seeing throughout. 
Spectra were obtained through a 1\arcsec\ longslit and dispersed with the 
R400 grating, yielding a resolution of $R\sim1000$. The grating was centered 
at 7400~\AA\ and the OG515 blocking filter was used; this setup provides
coverage from 5320~\AA\ to 9500~\AA. The e2V detectors were binned by a factor
of two in both the spatial and spectral directions, resulting in a spatial
scale of 0.15~arcsec~{pix}$^{-1}$ and a dispersion of 1.34~\AA~{pix}$^{-1}$.
Each target was observed with a single 1200s integration; based on an 
assessment of the 2D spectrum additional exposures were sometimes obtained 
to increase the $S/N$. No dithering in either the spatial or spectral 
directions was performed between successive exposures.
We observed a total of 17 targets, including all four of the 
likelihood candidates.

Data were processed in a standard fashion using the IRAF \texttt{gemini.gmos}
package. After bias subtraction and flat field correction, cosmic rays were
identified and masked using the LACOS \citep{lacos} routines. The 2D images 
were then stacked using \texttt{gemcombine}. The targets are faint and
the per pixel $S/N$ is quite low; we thus extracted 1D spectra by following a 
reference trace obtained from observations of bright quasars taken each night 
as part of another program. Wavelength calibration was provided by CuAr lamps. 
The spectra were flux-calibrated using observations of the spectrophotometric 
standards Wolf 1346 and Hiltner 600 obtained once per night.

The Gemini observations resulted in five newly confirmed quasars at
$4.5 < z < 5.4$ with a typical magnitude of $i=23.5$. The final spectra are 
displayed in Figure~\ref{fig:geminispec}. The Ly$\alpha$ line is readily 
apparent in each case. The 12 non-quasars present only featureless continua; 
they could be either late-type stars or faint red galaxies. An example 
spectrum of a failed target (non-quasar) is given in 
Fig.~\ref{fig:geminispec}. \citet{Matsuoka+17} have presented observations
of $z\sim6$ quasar candidates at a similar luminosity selected from Subaru
Hyper Suprime-Cam (HSC) imaging, and find a large number of sources with
narrow Ly$\alpha$ emission and no obvious quasar emission lines, or with
galaxy-like spectra. Some of our failed quasar targets may indeed be 
absorption line galaxies with no strong emission features, but the spectra
are of insufficient quality to establish their redshift and type.

\subsection{MMT}\label{sec:mmt}

We observed a large number of candidates with the 6.5m MMT using the Red 
Channel spectrograph during a number of observing runs from 2012 to 2014.
The MMT targets were primarily selected as backup targets for poor observing
conditions and were limited to $i<23$. One run from 2012 May 27-28 was 
dedicated to this program and conditions were excellent during this run
with 0{\farcs}7 seeing; we utilized this time for fainter targets 
($i \sim 23$).

We observed quasar candidates with the low dispersion 270~mm$^{-1}$ grating, 
typically centered at 7500~\AA\ with coverage from 5500\AA\ to 9700\AA. We 
alternated between the 1\arcsec\ or 1.5\arcsec\ slit based on the seeing, 
providing resolutions of $R\sim640$ and $R\sim430$, respectively.

Data processing employed standard longslit reduction methods using scripts
written in Python and using Pyraf\footnote{Pyraf is a product of the 
Space Telescope Science Institute, which is operated by AURA for NASA.}
routines. Basic corrections included bias subtraction, pixel level flat 
fields generated from internal lamps, and sky subtraction using 
a polynomial background fit along the slit direction. Cosmic rays were
identified and masked using the LACOS routines \citep{lacos}. Initial 
wavelength solutions were obtained from an internal HeNeAr lamp and then 
corrected using night sky lines (primarily the OH line list given by 
\citealt{Rousselot+00}); the final RMS for sky lines is $\sim2$\AA. 
Spectrophotometric standard stars were observed each night and used for 
flux calibration. However, the conditions were generally variable and the 
absolute flux calibrations are only approximate.

The MMT sample also includes five targets from Stripe 82 that did not
have spectroscopy at the time Paper I was prepared. These objects are
listed in Table~\ref{tab:notcanstab} in the Appendix.

Several of the targets observed with MMT have unusual properties. Two
quasars, CFHTLS J022112.3-034231 and CFHTLS J022112.6-034252, form a 
small-separation (20\arcsec) binary quasar at $z=5.02$; constraints
on quasar clustering at $z\sim5$ derived from this pair are discussed
in \citet{McGreer+16}. Another target, CFHTLS J141446.8+544631, is in
our final quasar candidate sample but is a $z=5.42$ lensed galaxy with 
strong Ly$\alpha$ emission. Additional spectroscopy and multiwavelength
observations of this object are presented in \citet{McGreer+17}.
Finally, CFHTLS J141956.4+555316 is tentatively assigned a redshift of
$z=5.0$ based on a marginal line detection at $\sim7310$\AA. The line
appears in multiple sky-subtracted 2D spectra, but extraction is hampered
by a strong complex of OH airglow lines at these wavelengths. This
object may be a weaker version of CFHTLS J141446.8+544631 but needs 
deeper spectroscopy to confirm its nature.

\subsection{LBT}\label{sec:lbt}

We obtained spectroscopy with the Large Binocular Telescope (LBT) 
Multi-Object Double Spectrograph (MODS1) instrument on 
2012 September 21. MODS provides moderate resolution optical 
spectroscopy \citep{Pogge+06}. The primary target was a Stripe 82 quasar 
from Paper I that has a close companion galaxy; however, we also observed 
five CFHTLS-W4 candidates based on an early version of the target 
selection. Details of the observations can be found in 
\citet{McGreer+14}; briefly, we used a 1\arcsec\ slit in good seeing
conditions ($\sim$0.8\arcsec) with the G670L grating ($R\sim1400$),
integrating for $\sim30$min on each target.

Only two of the five targets are $z\sim5$ quasars, and they are the only
two objects which appear in the final candidate list.

\begin{figure*}
 \epsscale{1.1}
 \plotone{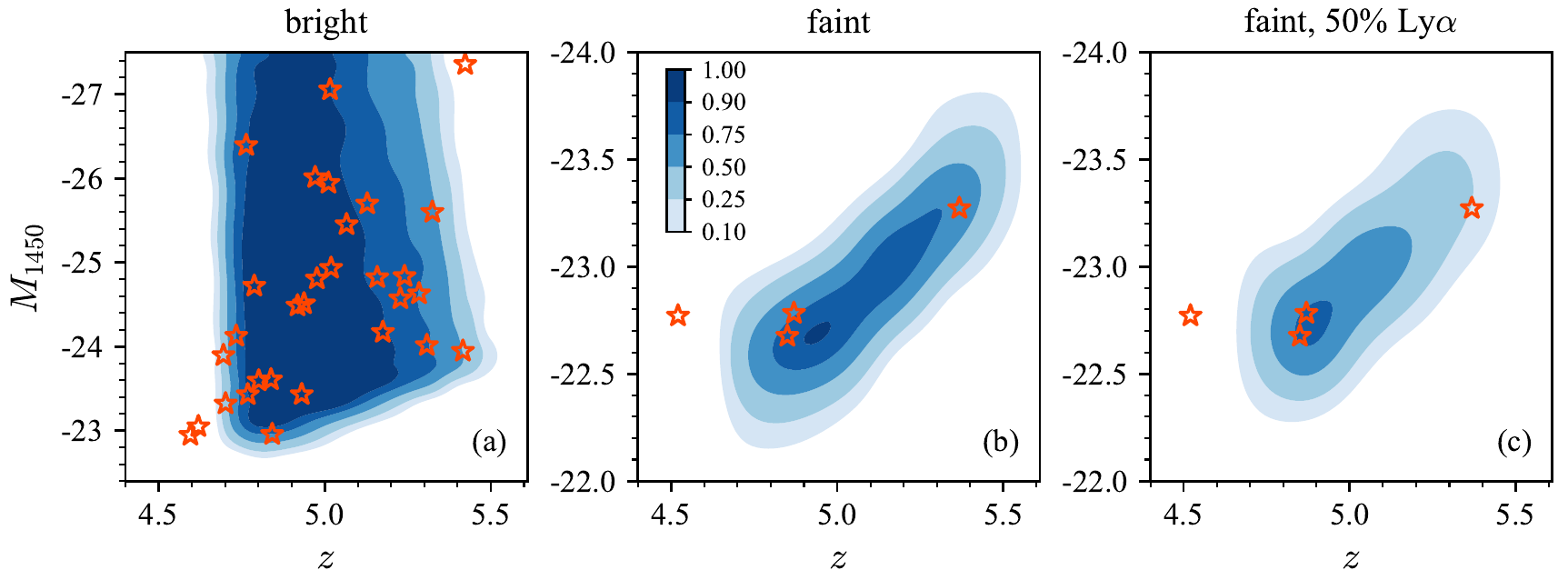}
 \caption{Selection functions for the color selection criteria 
 (\S\ref{sec:colorsel}). The filled contours represent the selection 
 efficiency with the scale given in panel (b). The completeness for 
 the bright selection ($i_{\rm AB}<23.2$) is presented in panel (a),
 and for faint selection ($23.2<i_{\rm AB}<23.7$) in panel (b). Panel (c)
 presents results for the faint criteria applied to a quasar model with
 the assumed Ly$\alpha$ EWs reduced by half. The dependence of the selection
 on Ly$\alpha$ strength is further discussed in the Appendix.
 Spectroscopically confirmed quasars from the final candidate list are
 marked with red stars.
 The $z=4.5$ faint quasar that lies well outside the selection region is 
 not formally likelihood-selected or color-selected, but observed because it 
 has $\pqsomidz>0.01$.
 \label{fig:selectionfun}
 }
\end{figure*}

\subsection{Redshift determination}

The spectra typically have modest $S/N$ ($\sim$few per pixel in the 
rest-UV continuum) and often the only well-detected feature is the
Ly$\alpha$ emission line and the continuum break due to the onset of
the Ly$\alpha$ forest. We obtain redshifts by fitting a set of 
templates to the spectra and varying the template redshifts, finding
the best-fit through a $\chi^2$ minimization. The set of templates
includes a fiducial quasar with emission line strengths and average
forest absorption as generated by our models (\S\ref{sec:simqso})
for a quasar with $M_{1450}=-24$. We also modify this fiducial template
by reducing the more prominent UV emission lines by a factor of two,
and add two narrow-line templates for which the broad components of
the UV lines are set to zero and the narrow Ly$\alpha$ and \ion{C}{4}
lines are either twice or half their nominal values from the models. 
Finally, we add a continuum-only template to represent a weak-lined 
quasar. Given that the templates are mainly fitting the Ly$\alpha$ 
feature the systematic uncertainty is $\Delta{z} \sim 0.1$.

The $k$-corrections are also determined from the quasar models, 
accounting for luminosity dependence (the Baldwin Effect) as in
Paper I. The $i$-band magnitudes and redshifts are matched to
model quasar spectra and used to determine the rest-UV continuum
luminosity $M_{1450}$. Dust extinction is not included in these
models.

\section{Results}

We obtained a total of 80 spectra of objects in the CFHTLS fields,
with 17 spectra from Gemini, 61 from MMT, and 2 from LBT. Redshifts for
two additional sources come from the BOSS DR9 quasar catalog \citep{dr9qso}.
The complete sample of bright candidates includes 97 targets in W1+W3+W4 
with $i<23.2$ (Table~\ref{tab:specbright}), of which over a third (39) have 
spectroscopic observations. This includes 38/87 (44\%) of the
color-selected candidates and 24/55 (44\%) of the likelihood-selected
candidates. The efficiency is quite high: 35/39 of the targets are $z>4$
quasars. For the faint W1 sample (Table~\ref{tab:specfaint}), 6/7 
were observed with Gemini/GMOS, including all of the likelihood-selected 
candidates (5), and 4/6 are $z\geq4.6$ quasars.

\subsection{Completeness}

\subsubsection{Color models}

We determine the completeness of our selection criteria by using the
color simulations to generate model quasar colors. The fraction of 
simulated quasars passing our color cuts is then calculated as a 
function of redshift and luminosity. This fraction is measured in a 
grid with bin widths ($\Delta{M}_{1450}$,~$\Delta{z}$) = (0.1,~0.05)
and with 100 quasars in each bin. Only the color criteria are considered
this calculation. Although the likelihood method was used to prioritize 
candidates, the color and likelihood methods have an equal amount of 
spectroscopic coverage. The color-selected sample also provides a more
well defined boundary for the selection cuts, especially considering 
that the same quasar model we use for the completeness calculation was 
used to train the likelihood model.

Figure~\ref{fig:selectionfun} displays the selection function for the
bright and faint quasar samples in the CFHTLS Wide survey. We derive
photometry from our quasar models by including a representation of
the CFHT photometric system with flux errors that match the depth of 
the Wide survey fields.
We also update the completeness calculation of the SDSS DR7 and 
Stripe 82 quasar samples from Paper I. The new calculation fixes an
error in $k$-correction used in Paper I that resulted in a shift of 
$\sim0.1$ mag in the absolute magnitudes.

For consistency, the quasar color models used here (\S\ref{sec:simqso}) 
are identical to those from Paper I and from the calculation of the 
luminosity function of BOSS quasars at $2<z<3.5$ \citep{Ross+13}. 
In Paper I we considered the effect of the Ly$\alpha$ emission on
$z\sim5$ quasar selection, noting that due to the Baldwin Effect we
expect fainter quasars to have stronger line emission and thus be
more easily selected by their redder $r-i$ colors (Ly$\alpha$ is in 
the $i$-band). However, if the properties of $z\sim5$ quasars are
different than the $z\sim3$ quasars from BOSS used to calibrate our
model, this would affect our completeness estimation. Panel (c) of
Figure~\ref{fig:selectionfun} compares the selection function derived
for a model where the Ly$\alpha$ flux is decreased by a factor of
two compared to the reference model (Panel b). Although the efficiency
is reduced, particularly at $z\ga5$, the overall effect is rather
modest and thus would not substantially alter the luminosity function
results. In the Appendix (\S\ref{sec:appendix}) we explore the dependence 
of the selection function on the assumed Ly$\alpha$ emission in 
greater detail and conclude that, although our color selection is less
sensitive to quasars with weak line emission, it is unlikely we are
missing a substantial population at $z\sim5$.

\begin{figure}
 \epsscale{1.1}
 \plotone{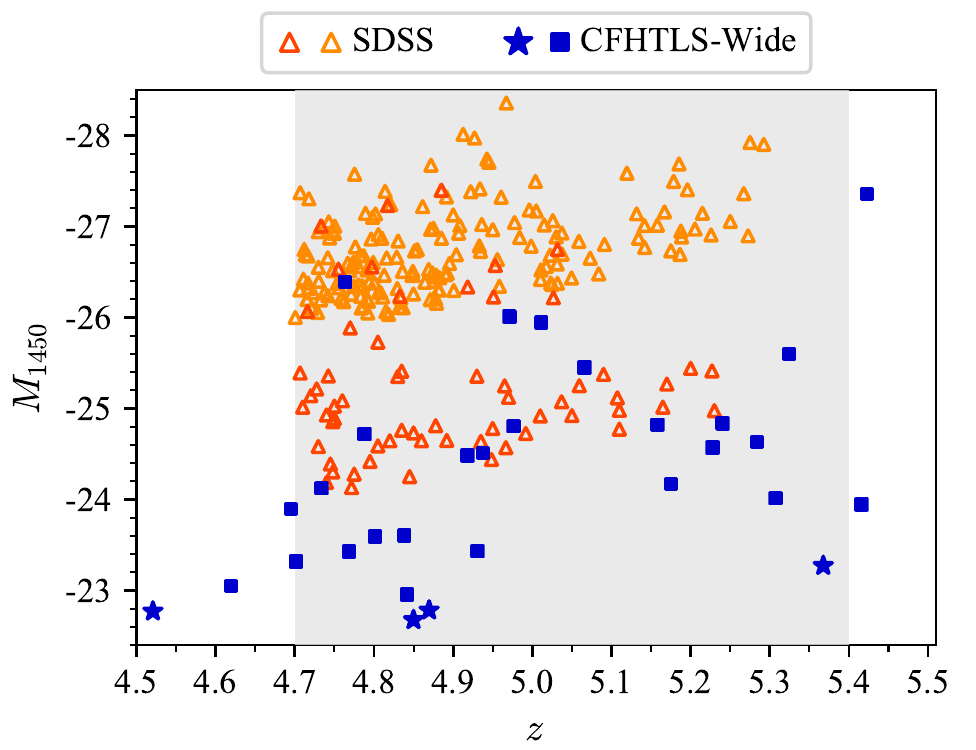}
 \caption{Distribution of the quasar samples in absolute UV
 magnitude and redshift. The updated results from SDSS DR7 (orange)
 and Stripe 82 (dark orange) are shown as empty triangles, truncated
 at $z>4.7$. The new CFHTLS quasars from the bright (blue squares) and 
 faint (blue stars) candidate samples overlap with the Stripe 82 results
 from Paper I, but reach $\sim1.5$ mag deeper.
 \label{fig:mz}
 }
\end{figure}

\subsubsection{Results from spectroscopy}

The results from the Gemini spectroscopy provide validation of the 
likelihood method as applied to the faint candidates: all four targets 
assigned the highest priority --- based on $\pqsomidz > 0.1$ --- were 
confirmed as $z>4.5$ quasars. Of the three additional targets selected 
with the strict color cuts but not likelihood, two were observed and 
neither were quasars.

In addition to the primary sample of seven color+likelihood targets, the
favorable observing conditions permitted observations of lower priority
targets just outside of the color selection boundary. Only one of these 
11 targets is a quasar: J020541.5-035350 has $i=23.16$ and is included
in the bright sample. With a redshift of $z=4.6$ it is just below the
target redshift range, which agrees with the crude photo$-z$'s provided
by likelihood: it has $\pqsomidz = 0.06$ and $\pqsolowz = 0.91$. 
The highest redshift quasar in the complete sample at $z=5.38$ has 
$\pqsomidz=0.10$, just at the threshold, but $\pqsohiz = 0.90$.

Many of the spectra for the bright candidates ($i<23.2$) were obtained 
before our selection criteria were finalized and thus probe regions 
outside of our selection boundaries. Of the 40 bright objects with 
spectroscopy not included in the final sample, 30 are not quasars and 
only eight are in the range $4.7 < z < 5.4$. Of those eight, five are
relatively bright quasars ($i_{\rm AB} \la 21$) that fail the rather
stringent ``dropout'' criteria in the $u$ and $g$ bands 
(eqns.~\ref{eqn:strict_usnr}\ and\ \ref{eqn:strict_gsnr}), and the
other three are just outside the color cuts. These three objects have
redshifts near the edges of our bin ($z=4.701, 5.228,~{\rm and}~5.364$)
where the selection efficiency is lower. This provides further confidence
in the completeness of our final selection criteria. The likelihood cut
also misses eight quasars in the mid-$z$ range 
(see Fig.~\ref{fig:likelihoods}), but again most of the missed quasars
are near the edge of the redshift bin.

\subsection{Photometric and Spectroscopic Completeness}

At the depths probed by our survey the CFHTLS imaging is highly complete.
By comparing the Wide survey catalogs to the overlapping Deep regions, we
find that the Wide imaging recovers $\sim96$\% of stellar sources in the
Deep best-seeing stacks at $i_{\rm AB}<23.7$, where most of the losses are
due to blending with nearby sources. Additionally, as discussed in
\S\ref{sec:stargal}, our star/galaxy separation method is $\sim90$\%
($\sim98$\%) complete to point sources for the faint (bright) sample. 
Thus we estimate the photometric completeness to be 86\% for the faint
sample and 94\% for the bright sample.

For the bright ($i_{\rm AB}<23.2$) quasar selection, we have obtained
spectra for all targets with $i_{\rm AB}\le20.8$, 86\% of the targets
with $20.8<i_{\rm AB}\le22.6$, and 37\% of targets with 
$22.6<i_{\rm AB}\le23.2$. For the faint quasar selection we have Gemini
spectra for 6/7 targets. Both the photometric and spectroscopic
completenesses are applied during the calculation of the luminosity 
function.

\subsection{Binned Luminosity Function}

The binned QLF is determined separately for three samples: ``SDSS Main''
consists of quasars from SDSS DR7. There are slight differences between
the results presented here and those from Paper I due to the updated
$K$-corrections and completeness models; otherwise, the observed sample
of quasars is identical. ``SDSS Stripe 82'' is similarly updated from
Paper I, but includes the five additional quasars with spectroscopic 
observations presented here. Finally, ``CFHTLS Wide'' consists
of the results from the spectroscopic observations of targets in the
CFHTLS Wide fields, including both the MMT, LBT, and Gemini observations of 
$i<23.2$ targets and the Gemini observations of $23.2<i<23.7$ targets.
Although the selection methods for each of the three samples are
slightly different, the quasar models and methodology used to calculate 
the completeness corrections are identical.

The nominal CFHTLS selection function extends to slightly higher 
redshifts than the SDSS and Stripe 82 samples from Paper I. We 
thus adopt a wider redshift bin of $4.7 < z < 5.4$ for the CFHTLS data, 
compared to the $4.7 < z < 5.1$ bin that was used in Paper I. However,
we continue use the smaller bin for the recalculation of the SDSS and
Stripe 82 QLF. The choice of a larger bin includes more of the quasars
found in CFHTLS, but has little effect on the luminosity function as the
number density declines steeply with redshift and the mean quasar
redshift is $\langle{z}\rangle \approx 4.9$ for all three samples.

\begin{figure}
 \epsscale{1.1}
 \plotone{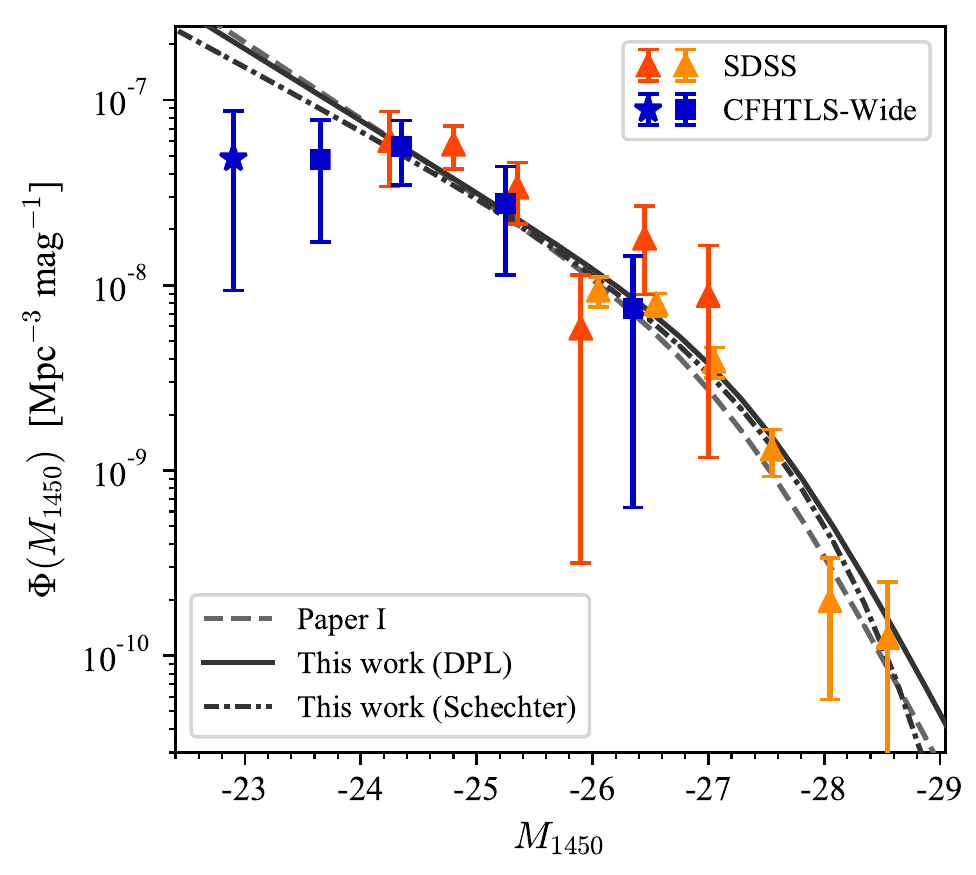}
 \caption{Binned QLF results at $z\sim5$ shown as points with error bars,
 including the SDSS Main (orange), SDSS Stripe 82 (dark orange), 
 and CFHTLS-Wide (blue) samples. The lines denote model fits to the QLF
 using maximum likelihood. The dashed line is the best-fit from Paper I
 using SDSS data only, the solid line is the best-fit double power law model
 from this work, and the dot-dashed line the best-fit Schechter function
 model. The Stripe 82 and CFHTLS points show good agreement where they 
 overlap in luminosity. The faintest bin from the CFHTLS, based on the W1
 Gemini spectroscopy and represented by a star, is well below the best fit 
 QLF models.
 \label{fig:qlfcompare}
 }
\end{figure}

The results of the binned QLF calculation are given in 
Table~\ref{tab:binnedqlf}. To calculate $\sigma_{\Phi}$ we have 
adopted the approximation for confidence intervals in the low count
regime provided by \citet{Gehrels86}. Figure~\ref{fig:qlfcompare} presents
our measurement of the QLF over a dynamic range in luminosity greater than 
a factor of 100, to the limit of $M_{1450}=-22.9$ achieved with the faint
CFHTLS quasars.

The results from the CFHTLS fields agree well with those from SDSS Main
and Stripe 82 in the luminosity range over which they overlap. Although 
the color selection methods used to target quasars in these surveys share
a high degree of similarity, the input imaging data is quite different,
providing some encouragement that the results are not sensitive to the
systematics of any one survey.

\begin{deluxetable}{rrrrr}
 \centering
 \tablecaption{Binned QLF\label{tab:binnedqlf}}
 \tablewidth{0pt}
 \tablehead{
  \colhead{$M_{1450}$} &
  \colhead{$N$} &
  \colhead{$N_{\rm cor}$} &
  \colhead{$\log\Phi$\tablenotemark{a}} &
  \colhead{$\sigma_\Phi$\tablenotemark{b}} 
 }
 \startdata
\multicolumn{5}{c}{SDSS Main} \\
-28.55 &  1 &   1.7 & -9.90 & 0.12 \\
-28.05 &  2 &   2.6 & -9.70 & 0.14 \\
-27.55 & 13 &  17.2 & -8.89 & 0.37 \\
-27.05 & 34 &  51.5 & -8.41 & 0.72 \\
-26.55 & 67 & 100.4 & -8.10 & 1.08 \\
-26.05 & 30 &  38.6 & -8.03 & 1.74 \\[4pt]
\multicolumn{5}{c}{SDSS Stripe 82} \\
-27.00 &  3 &   4.9 & -8.06 & 5.57 \\
-26.45 &  7 &   9.9 & -7.75 & 6.97 \\
-25.90 &  3 &   3.2 & -8.23 & 3.38 \\
-25.35 & 12 &  18.6 & -7.47 & 10.39 \\
-24.80 & 20 &  31.6 & -7.24 & 13.12 \\
-24.25 &  8 &  14.9 & -7.22 & 21.91 \\[4pt]
\multicolumn{5}{c}{CFHTLS Wide} \\
-26.35 &  3 &   3.5 & -8.12 & 4.34 \\
-25.25 &  5 &   8.8 & -7.56 & 12.70 \\
-24.35 & 10 &  18.0 & -7.25 & 18.05 \\
-23.65 &  4 &   7.8 & -7.32 & 23.77 \\
-22.90 &  3 &   5.1 & -7.32 & 28.24 \\
 \enddata

 \tablenotetext{a}{$\Phi$ is in units of Mpc$^{-3}$~mag$^{-1}$.}
 \tablenotetext{b}{$\sigma_\Phi$ is in units of $10^{-9}$~Mpc$^{-3}$~mag$^{-1}$.}
\end{deluxetable}

The most striking result apparent in Figure~\ref{fig:qlfcompare}
is the low number of quasars in the faintest luminosity bin. The 
space density of quasars at $M_{1450}=-22.9$ is roughly similar 
to that at  $M_{1450}=-23.7$. This suggests that the relatively 
steep faint-end slope determined in Paper I  ($\alpha$=-2.03) may not 
extend to lower luminosities. We now explore parametric model fits to 
the new QLF data.

\begin{figure*}
 \epsscale{1.1}
 \plotone{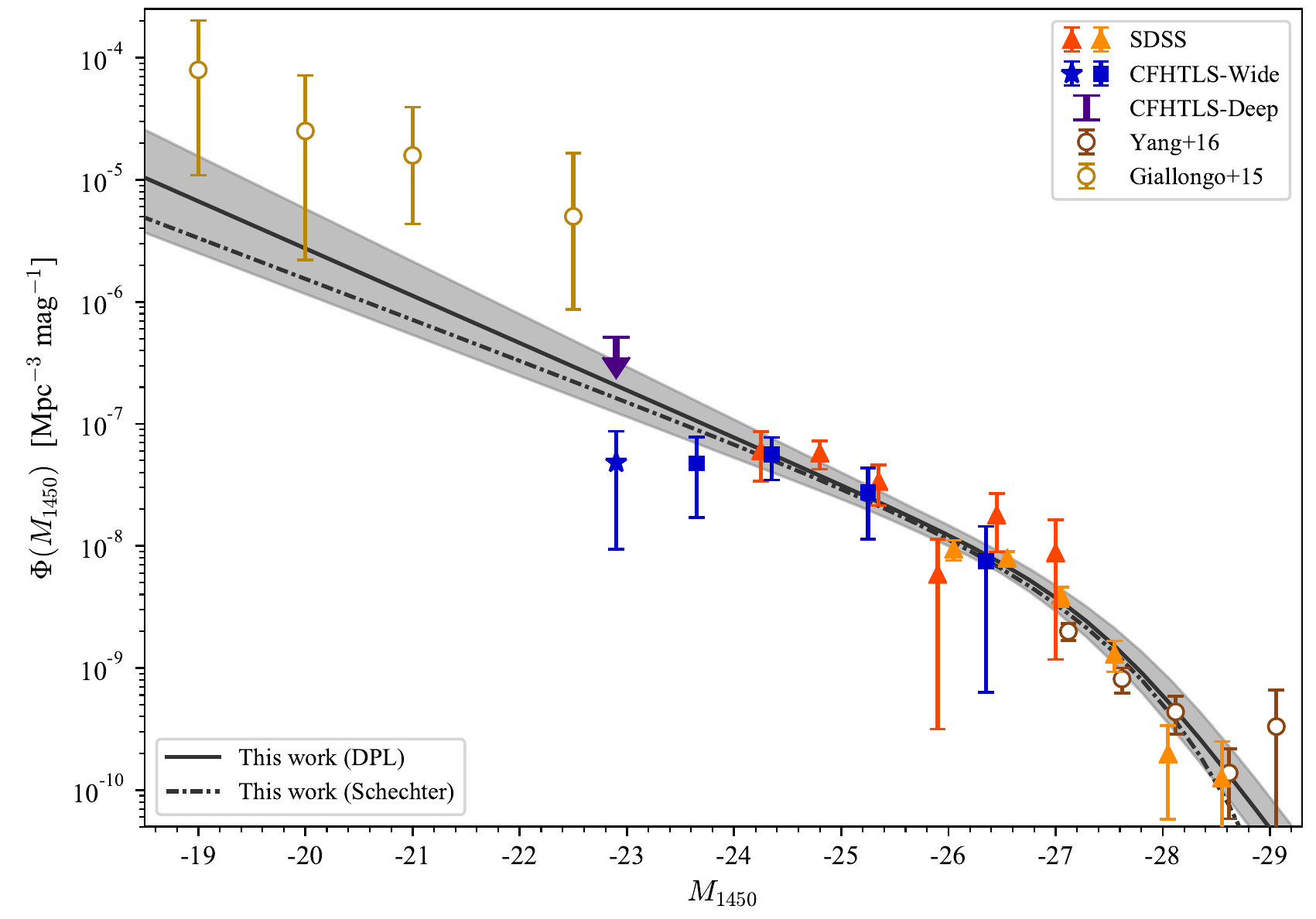}
 \caption{QLF from this work compared to \citet{Yang+16} (brown) and
 \citet{Giallongo+15} (dark yellow) at the bright and faint ends, respectively.
 The purple downward arrow marks the $3\sigma$ upper limit obtained from
 the CFHTLS Deep D1+DXS region. The remaining lines and points are as in
 Figure~\ref{fig:qlfcompare}, with the addition of a gray boundary spanning
 the $3\sigma$ range allowed by the uncertainties in the double power law fits.
 Although the CFHTLS sample does not overlap the \citet{Giallongo+15} sample
 in luminosity, it is clear that the two QLFs do not agree at the
 faint end.
 \label{fig:qlfcompare2}
 }
\end{figure*}

\subsection{Parameter Estimation from Maximum Likelihood}

As in Paper I, we employ maximum likelihood estimation to obtain
parametric model fits to our data. Assuming an evolving luminosity
function $\Phi(M,z)$ modulated by a selection function $p(M,z)$, 
the log likelihood function is 
\[
	S = -2 \sum\limits_i^N \ln[\Phi(M_i,z_i)] 
	      + 2\int\int\Phi(M,z)p(M,z)\frac{dV}{dz} dM dz ~.
\]
We first fit a double power law function,
\[
	\Phi(M,z) = \frac{\Phi^*}
	     {10^{0.4(\alpha+1)(M-M^*)} + 10^{0.4(\beta+1)(M-M^*)}} ~,
\]
where $\Phi^*$ is the normalization of the space density, $M^*$ is 
the break luminosity, and $\alpha$ and $\beta$ are the faint- and
bright-end slopes, respectively. We consider evolution of this function
within our bin of width $\Delta{z}=0.7$ by allowing the normalization
to decline as a power law, $\Phi^*(z) = \Phi^*(z=6)\times10^{k(z-6)}$, 
with $k=-0.47$ \citep{Fan+01LF}. In Paper I we examined the evolution
of both $\Phi^*$ and $M^*$ at $z \ge 4$ and found that $k=-0.47$ provides
a good fit to the data out to $z\sim5$, although there is some indication
this value steepens at $z\ga6$. $M^*$ is also found to evolve strongly
out to $z\sim5$, although its continued evolution to higher redshift is
difficult to assess and here we ignore any evolution in this parameter
within our bin.
We experimented with fits using $k=-0.7$ and find that the results within
our redshift bin are generally insensitive to the choice of the $k$ term.

The apparent flattening of the faint-end slope based on the new
measurements presented here introduces some tension with the
double power law form. We thus experiment with a Schechter function
parameterization of the QLF:
\[
	\Phi(M,z) = \frac{\ln{10}}{2.5}\Phi^*
	     10^{0.4(\alpha+1)(M-M^*)}\exp[-10^{0.4(M-M^*)}] ~.
\]
This model requires only three parameters, similar in nature to the
double power law parameters except the power-law bright-end slope is
replaced by an exponential cutoff.
While this functional form is commonly used to describe the galaxy
luminosity function, a double power law is generally preferred by
QLF measurements.

Table~\ref{tab:mlefit} contains the results from the maximum likelihood 
parameter estimation. We include the fit results from Paper I in the 
second column for comparison, updated to the cosmology adopted in this paper.
In the third and fourth columns we present the best-fit results for the 
double power law and Schechter luminosity function parameterizations, 
respectively. Following Paper I, we have fixed the bright-end slope for the
double power law to $\beta=-4$, thus the two functional forms have the same
number of parameters. The log-likelihood values differ at the $<1\sigma$ 
level, indicating that the two forms provide an equally good fit.

\begin{deluxetable}{rccc}
 \centering
 \tablecaption{MLE fit parameters\label{tab:mlefit}}
 \tablewidth{0pt}
 \tablehead{
  \colhead{Parameter} & \colhead{Paper I (DPL)} & \colhead{DPL} & \colhead{Schechter}
 }
 \startdata
           $\log\Phi^*$\tablenotemark{a} & $-8.94^{+0.20}_{-0.24}$ & $ -8.97^{+0.15}_{-0.18}$  & $ -8.70^{+0.19}_{-0.22}$ \\
                            $M_{1450}^*$ & $-27.21^{+0.27}_{-0.33}$ & $-27.47^{+0.22}_{-0.26}$  & $-27.33^{+0.26}_{-0.32}$ \\
                                $\alpha$ & $-2.03^{+0.15}_{-0.14}$ & $ -1.97^{+0.09}_{-0.09}$  & $ -1.84^{+0.12}_{-0.11}$ \\
                                 $\beta$ & $-4.00$ &  $-4.00$  &  - \\
 \enddata
 \tablecomments{Parameters without uncertainty ranges are fixed during the 
maximum likelihood fitting.}

 \tablenotetext{a}{$\log\Phi^* \equiv \log\Phi^*(z=6) + k(z-6)$, with $k=-0.47$.}
\end{deluxetable}

\subsection{Comparison to Previous Work}

We first compare our results to those from Paper I. Comparison of the 
best-fit double power law values in Table~\ref{tab:mlefit} shows that
the new fits are in good agreement with the previous results, with
all parameters agreeing within the $1\sigma$ ranges. The QLF models shown 
in Figure~\ref{fig:qlfcompare} demonstrate that the new double power law
fit has almost no effect on the faint end number counts, although the 
Schechter form predicts number counts at lower luminosities that are 
a factor of $\sim2$ lower than the double power law extrapolation.

Figure~\ref{fig:qlfcompare2} places our new QLF measurement in the context
of other measurements at $z\sim5$. 
The bright end of the luminosity function was recently assessed
by \citet{Yang+16} using a combination of SDSS and WISE colors. At
$M_{1450} \le -28$ our results are in good agreement. Our number densities
are slightly higher at lower luminosities, which may be due to the 
efficiency of WISE selection at these redshifts. \citet{Yang+16} found a 
best-fit bright-end slope of $\beta=-3.6$, somewhat flatter than the fixed 
value of $\beta=-4$ we adopted; however, our results are rather insensitive 
to the value of $\beta$. They also find a somewhat fainter break luminosity 
of $M_{1450}^* = -27$, although the two results agree within the $1\sigma$
uncertainties.

At the faint end of the luminosity function we find a much lower number
density than implied by the results of \citet{Giallongo+15} based on 
photometric redshifts of putative X-ray detections in the GOOD-S field. 
Where the two measurements nearly overlap our counts are lower by more
than one order of magnitude. We consider the maximum possible number 
density consistent with our survey in two ways. First, we assess the 
allowed range of QLF fits by performing 1000 Monte Carlo samplings of the 
best-fit double power law parameters, keeping those with a log-likelihood 
within $1\sigma$ of the best-fit result. This range is marked by the gray 
shaded region. Second, we obtain a density from the single quasar in the 
D1 field. The $3\sigma$ upper limit for the density at $M_{1450}=-22.9$ 
obtained from this object is denoted by the purple error bar with a 
downward arrow. As argued in \S\ref{sec:d1dxs}, we expect the selection 
in the D1 region to be highly complete and thus this upper limit provides 
a strong constraint on the faint number counts. Recently, \citet{PDM17} 
performed a re-analysis of the \citet{Giallongo+15} sample, and find a 
number density at $z\sim5$ about a factor of $\sim3$ lower, in much 
better agreement with our results. 

Repeating the procedure of \citet{Giallongo+15}, we derive an ionizing 
emissivity at $z\sim5$ that is almost an order of magnitude lower: 
$\epsilon_{912}^{24} = 0.8$ as opposed to 
$\epsilon_{912}^{24} = 5.9$ in \citet{Giallongo+15} (compare to
$\epsilon_{912}^{24} = 1.3$ in \citealt{PDM17}).
Based on these results it is highly unlikely that faint AGN make a 
significant contribution to hydrogen reionization 
\citep[see also][]{DAloisio+17,Khaire17,Ricci+17}, 
unless our survey is highly incomplete or the extrapolation to fainter 
sources does not follow the QLF we have derived.

Finally, we compare our results to measurements of the X-ray QLF from 
wide-area surveys as given in 
\citet[][see also \citealt{Marchesi+16}]{Georgakakis+15}. Assuming typical
values for the X-ray/optical flux ratio, \citet{Georgakakis+15} found
that the QLF we obtained in Paper I is nearly an order of magnitude below
the X-ray counts at a similar luminosity. Our new results favor an even
flatter slope at the faint end. This suggests that a significant fraction
of black hole growth at low luminosities may be highly obscured 
\citep[see also][]{Trakhtenbrot+16}.

\subsection{Conclusions}

We present a sample of 104 candidate $z\sim5$ quasars drawn from the 
CFHTLS Wide survey. Spectroscopic confirmations for 37 quasars were
obtained using Gemini-GMOS, MMT Red Channel, and LBT-MODS. The 
luminosity function derived from these quasars is in good agreement
with our previous measurements using SDSS Stripe 82 (Paper I). The
faintest quasars in the sample reach $M_{1450}=-22.9$, with the 
full luminosity function extending over a range of 6~mag.

Parametric fits to the luminosity function obtained using a maximum
likelihood method show that the break in the luminosity function
is at $M_{1450} \approx -27$, in agreement with the evolutionary model
presented in Paper I that includes a steady increase in the break 
luminosity with increasing redshift. Although the best-fit faint-end
slope is somewhat steep ($\alpha \approx -2$), the data in the lowest
luminosity bins are below the best-fit QLF, suggesting the faint number
counts may be even fewer. In addition to the traditional double power-law
form for the QLF, a Schechter function is found to provide an equally
good fit to the overall data, and provides a marginally better fit to
the faint number counts.

These results do not support a scenario in which faint AGN provide a
significant contribution to the hydrogen-ionizing radiative background
at $z\sim5$. Integrating the best-fit QLF to $M_{1450}=-18$ results in
quasars producing only a few per cent of the ionizing photons required
to maintain hydrogen ionization. This is contrary to recent claims of
a significant faint AGN population based on photometric redshifts
of galaxies associated with X-ray emission 
\citep[][although see \citealt{PDM17}]{Giallongo+15}.

We consider potential sources of incompleteness in our survey, focusing
on the effect of Ly$\alpha$ emission as this has a significant impact
on optical color selection. We find marginal evidence for weaker Ly$\alpha$
emission in our faintest quasars compared to a luminosity-matched
sample at $z\sim3$ from BOSS. If Ly$\alpha$ is systematically weaker at 
high redshift, our selection efficiency may be overestimated. However, 
this is unlikely to substantially alter the conclusion that quasars are
insufficient in number to drive hydrogen reionization. Future studies 
that can efficiently select high-$z$ quasars using methods independent
of the Ly$\alpha$ flux -- e.g., color selection in infrared bands 
\citep{Wang+16,Yang+16}, or variability selection 
\citep{Peters+15,AlSayyad16} -- will address this issue and form a 
more complete picture of the faint quasar population near the epoch
of reionization.

\section{Acknowledgments}

I. D. McGreer and X. Fan acknowledge the support from the U.S. NSF grant 
AST 15-15115.
L.J. acknowledge support from the National Key R\&D Program of China (2016YFA0400703).
The work presented here is based in part on observations obtained at the 
Gemini Observatory, which is operated by the Association of Universities for 
Research in Astronomy, Inc., under a cooperative agreement with the NSF on 
behalf of the Gemini partnership: the National Science Foundation (United 
States), the National Research Council (Canada), CONICYT (Chile), the 
Australian Research Council (Australia), Minist\'{e}rio da Ci\^{e}ncia, 
Tecnologia e Inova\c{c}\~{a}o (Brazil) and Ministerio de Ciencia, 
Tecnolog\'{i}a e Innovaci\'{o}n Productiva (Argentina). Gemini data were
processed using the Gemini IRAF package and \texttt{gemini\_python}.
Additional observations were obtained at the MMT Observatory, a joint facility 
of the Smithsonian Institution and the University of Arizona.
Also based on observations obtained with MegaPrime/MegaCam, a joint project of CFHT and CEA/DAPNIA, at the Canada-France-Hawaii Telescope (CFHT) which is operated by the National Research Council (NRC) of Canada, the Institut National des Science de l'Univers of the Centre National de la Recherche Scientifique (CNRS) of France, and the University of Hawaii. This work is based in part on data products produced at the Canadian Astronomy Data Centre as part of the Canada-France-Hawaii Telescope Legacy Survey, a collaborative project of NRC and CNRS.

This work made use of the following open source software: 
IPython \citep{ipython}, matplotlib \citep{matplotlib}, 
NumPy \citep{numpy}, SciPy \citep{scipy},  
Astropy \citep{astropy}, and astroML \citep{VanderPlas+12}.
This research has made use of NASA's Astrophysics Data System.

{\it Facilities:} 
 \facility{Gemini:Gillett (GMOS)}, 
 \facility{LBT (MODS1)},
 \facility{MMT (Red Channel spectrograph)}, 
 \facility{CFHT (MegaCam)}

\bibliographystyle{hapj}
\bibliography{z5cfhtls}

\section{Appendix}\label{sec:appendix}

\subsection{Color selection and weak-lined quasars}

The efficiency of our selection method relies on the colors of $z\sim5$ 
quasars becoming redder in $r-i$ and bluer in $i-z$ than stars in a 
similar region of color space~(Fig.~\ref{fig:simcolors}). This is due to 
a combination of the attenuation in bands blueward of the $i$-band from 
the strong IGM absorption, and the generally prominent 
Ly$\alpha$ emission from the quasar. Because Ly$\alpha$ is within the 
$i$-band at these redshifts, stronger Ly$\alpha$ emission increases the 
contrast between quasars and stars and hence the selection efficiency. We 
noted this effect in Paper I (Fig. 13 and surrounding discussion), where
we showed that our color selection is expected to be relatively 
{\em more} sensitive to fainter quasars, which tend to have stronger 
Ly$\alpha$ emission associated with the Baldwin Effect \citep{Baldwin77}.

A potential concern is the fraction of quasars with weak emission lines.
Using our color simulations, we find that removing the emission lines
completely (the most extreme case of a weak-lined quasar) indeed shifts
the $riz$ colors towards the boundary of our color box.
In Paper I, we noted that the fraction of weak-lined quasars, defined
as EW$_0$(\lyanv) $< 15$\AA\ by \citet{DS09}, is $\sim6\%$ for SDSS 
quasars at $z\ga4$ where the selection efficiency of weak-lined objects 
is relatively high (see their Fig. 5). We thus expect a minimal correction 
to the luminosity function from this class of objects unless the weak-lined
fraction evolves dramatically from $z\sim4$ to $z\sim5$, or if it is much
higher for the low luminosity objects in our sample compared to the more
luminous SDSS quasars in \citet{DS09}.

Weak emission lines may arise due to intrinsic properties of quasars, 
e.g., a softer ionizing continuum or shielding gas 
\citep{Wu+12,Plotkin+15,SL15}. While it is not known whether or
not these properties evolve with redshift, in general, the spectral
properties of quasars appear to show little redshift evolution
\citep[e.g.,][]{Kuhn+01,Yip+04,Jiang+06}. On the other hand, the 
Ly$\alpha$ emission may be affected by strong neutral absorbers near
the quasar redshift. While the line-of-sight proximity effect is
generally suppressed in lower redshift, high-luminosity quasars
\citep[][although see \citealt{DWW08}]{BDO88,Hennawi+06} likely due to 
the strong ionizing radiation from the quasar, in our sample of faint, 
$z\sim5$ quasars the incidence of self-shielded absorption systems may
be greater \citep[see related discussions in][]{HP07,BH13}, 
the mean UV background emission is weaker \citep{Calverley+11}, 
and the expected proximity zone sizes are smaller \citep[e.g.,][]{Eilers+17},
all of which may lead to greater Ly$\alpha$ absorption.
Additionally, if the Ly$\alpha$ emission line in high-$z$ quasars 
is systematically blueshifted it may be suppressed by strong IGM
absorption from dense gas, even if it is highly ionized
\cite[cf.][for a discussion of red damping wings from highly ionized environs]{Keating+15}.
Ly$\alpha$ blueshifts are correlated with \ion{C}{4} blueshifts \citep{KH09},
which are commonly found in luminous quasars 
\citep{Mortlock+11,Richards+11,BB15}.

In the most extreme case, if low-luminosity $z\sim5$ quasars tend to
have extremely weak Ly$\alpha$ emission, our selection efficiency would 
be greatly reduced. In this Appendix we examine in detail the dependence 
of our color selection efficiency on the Ly$\alpha$ emission properties. 
Our quasar models are calibrated using SDSS/BOSS quasars at $2<z<3.5$; 
any differences between the spectra of $z\sim3$ and $z\sim5$ quasars will 
impact our luminosity function calculation.

\subsection{Quasar samples with Ly$\alpha$ coverage}

Our comparison sample is constructed from $z\sim3$ quasars from the SDSS
DR7QSO catalog \citep{dr7qso} and the BOSS DR9QSO catalog \citep{dr9qso}.
For the DR7QSO quasars we adopt the redshifts provided by \citet{HW10},
and for DR9QSO we use the PCA redshifts (Z$_{\rm PCA}$, \citealp{dr9qso}).
We then select quasars with $2.4 < z < 3.5$, providing coverage of the
Ly$\alpha$ region. For the BOSS quasars we further restrict the sample
to quasars in Stripe 82, which are predominantly selected by variability
\citep{PD11}, reducing any color selection bias that would disfavor objects 
with weak lines. The BOSS sample includes $\sim900$ quasars that were
observed in SDSS I/II and present in the DR7QSO sample. The BOSS and SDSS
spectra are independent and we choose to retain these objects in both
samples. The final $z\sim3$ comparison set consists of $\sim$5700 BOSS and 
$\sim10000$ SDSS quasars and spans $-28 \la M_{1450} \la -23$.

We also draw on the DR7QSO and DR9QSO catalogs to select luminous $z\sim5$
quasars, obtaining $\sim400$ ($\sim150$) quasars from SDSS (BOSS) in the
range $4.5 < z < 5.5$. Finally, we include  fainter quasars from our own 
survey, with $\sim75$ MMT spectra and 5 Gemini spectra of $z\sim5$ quasars.

\begin{figure*}[!ht]
 \epsscale{1.1}
 \plotone{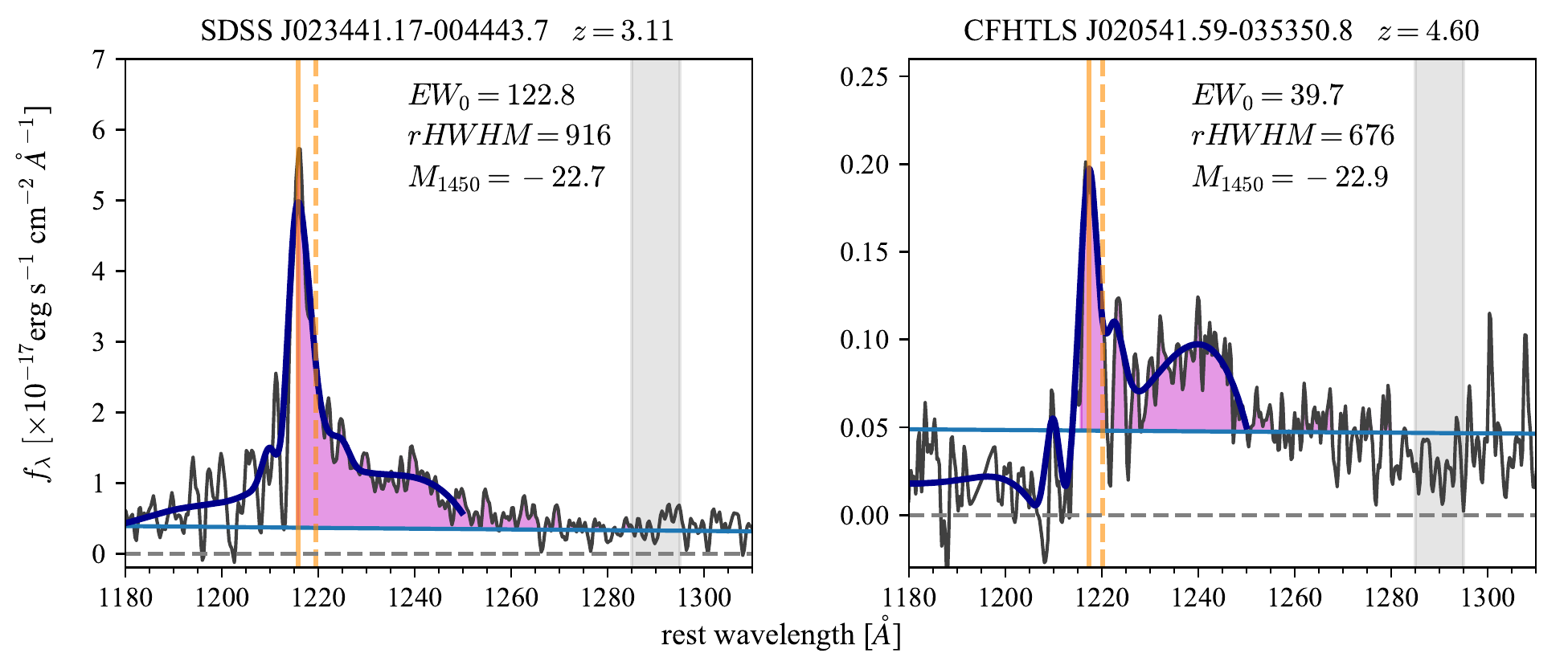}
 \caption{Examples of non-parametric estimation of line features. The
 left panel displays the spectrum of a $z\sim3$ quasar from BOSS with a 
 gray line, and the right panel a Gemini spectrum of a $z\sim5$ quasar 
 at a similar luminosity. The light blue line represents the power-law
 continuum fit as described in the text; the gray shaded region is the
 bluest spectral window used for continuum fitting. The dark blue line
 traces the smoothed spline model fit to the spectrum. The total EW of
 the Ly$\alpha$+\ion{N}{5} region is obtained by direct integration above
 the continuum fit, as represented by the purple shaded region. The 
 Ly$\alpha$ red HWHM is obtained from the peak wavelength (solid orange
 line) and the wavelength at which the smoothed profile drops to 60\%
 of the peak value (dashed orange line, scaled to HWHM position for a
 Gaussian profile).
 \label{fig:lyafitexamples}
 }
\end{figure*}

\subsection{Method}

The Ly$\alpha$ emission region of high-redshift quasars consists of a 
complex set of emission and absorption features. We initially attempted 
to fully model the Ly$\alpha$ region using a series of Gaussians for 
the Ly$\alpha$ $\lambda$1216, \ion{N}{5} $\lambda$1240, and 
\ion{Si}{2} $\lambda$1261 emission features, including both a broad and 
narrow component for Ly$\alpha$. However, we found that reliable 
decomposition of this region into multiple Gaussians was extremely 
difficult due to the varied shapes of the line profiles, strong absorption 
features (from both intrinsic Ly$\alpha$ and BAL-type absorption), 
Ly$\alpha$ forest absorption, and the relatively low $S/N$ of the spectra.

We thus adopted non-parametric approaches to characterize the Ly$\alpha$
emission. First, we quantify the {\em total} EW of the emission region
(including Ly$\alpha$, \ion{N}{5}, and \ion{Si}{2}) using the method of
\citet{DS09}. Briefly, a power-law continuum is fit to a set of
spectral windows relatively uncontaminated by emission lines. After
dividing the power-law continuum, the total EW is obtained by summing
the residual (positive) emission; as in \citet{DS09} we refer to this
quantity as EW$_0$(\lyanv). One difference with the method
of \citet{DS09} is that we sum over the wavelength range $1216 < \lambda < 1290$
instead of $1160< \lambda < 1290$. Truncating the blue edge of the region
at Ly$\alpha$ provides a more reasonable comparison between $z\sim3$ and
$z\sim5$ spectra, as otherwise the results would be affected by the increased
IGM absorption blueward of Ly$\alpha$ between those redshifts.

Next, we roughly quantify the width of the core the Ly$\alpha$ line, 
which is typically dominated by the narrow component. For this we use
the half-width at half-maximum of the profile redward of line
center (rHWHM). Using the red side avoids regions affected by Ly$\alpha$ 
forest absorption.
However, the red side is affected by blending from \ion{N}{5} emission
and also strong absorption features that appear in many spectra (usually
\ion{N}{5} BAL-type absorption). To address these issues, we fit a heavily 
smoothed spline model to the line profile. The spline models were derived 
by  iteratively fitting splines of increasing flexibility while masking 
pixels significantly below the low-order fits to remove absorption
features.

After obtaining the spline models, we calculate the half-width at 65\% 
of the peak, instead of 50\%, as we found this value to be less affected 
by contamination from the neighboring lines while still providing a 
reasonable estimate of the line width. The calculated value is then 
scaled appropriately for a Gaussian to the half-power width in order to
obtain the rHWHM estimate. Examples of Ly$\alpha$ fitting for a BOSS
quasar at $z=3.1$ and a Gemini spectrum of a faint CFHTLS quasar at
$z=4.6$ are presented in Figure~\ref{fig:lyafitexamples}.

We stress that the both quantities (EW$_0$ and rHWHM) are obtained using
exactly the same procedure on the spectra from the different quasar
samples.

\begin{figure*}[!ht]
 \epsscale{1.1}
 \plotone{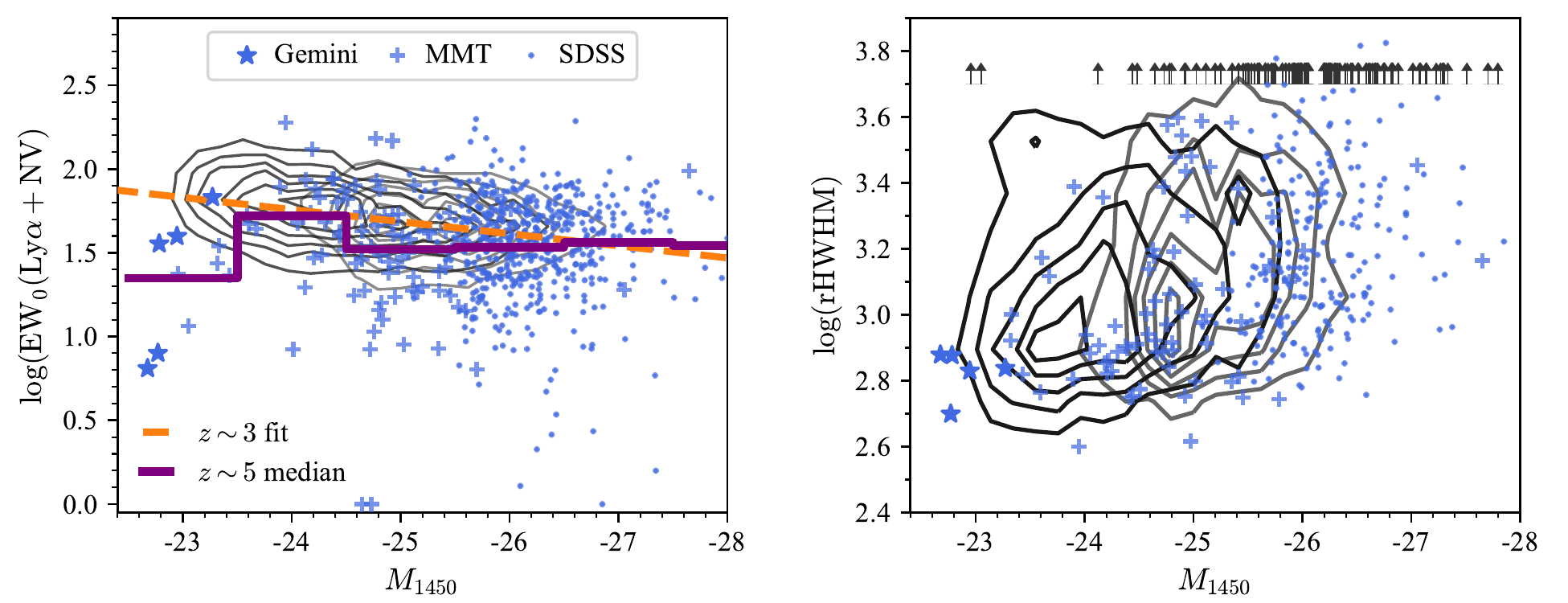}
 \caption{Left: Distribution of rest-frame EW of the combined 
 Ly$\alpha$+\ion{N}{5} emission lines for various high-redshift quasar 
 samples. The black (gray) contours represent the distribution of EWs 
 measured from luminous quasars at $2.4<z<3.5$ from DR7QSO (DR9QSO).
 The $z\sim5$ data are represented with circles. A running median of the
 $z\sim5$ points in bins of width $\Delta{m}=1$ is marked with the solid
 brown line.
 Right: rHWHM measurements. Lines and symbols are same as the left panel.
 Objects with rHWHM $> 5000~\kms$ are displayed as upward pointing arrows
 at $\log({\rm rHWHM}) = 3.7$; these cases are usually due to contamination
 from broad Ly$\alpha$ or \ion{N}{5}.
 \label{fig:lyafits}
 }
\end{figure*}

\subsection{Results}

Figure~\ref{fig:lyafits} displays measurements of EW$_0$(\lyanv) from our
quasar samples. The grayscale contours represent the distribution of EWs
measured from $z\sim3$ SDSS/BOSS quasars, and show a clear trend of 
increasing EW with decreasing luminosity (the Baldwin Effect). This trend
can be recovered with a linear fit to the data, represented by the dashed
orange line which is $\log(\EW) = 1.7 + 0.07(M_{1450}+25)$.
The scatter
points represent the measurements from $z\sim5$ quasars. At high luminosities
they overlap the $z\sim3$ data, but at the lowest luminosities there appears
to be a dearth of high-EW quasars as expected from the extrapolation of the
$z\sim3$ sample. The solid purple line marks the median EW in bins of
width 1~mag in luminosity. For $z\sim5$ quasars with a luminosity 
$M_{1450} \approx -23$ the median is 
$\langle$EW$_0$(\lyanv)$\rangle = 1.3$, which is $\sim0.5$~dex lower than
the relation for the $z\sim3$ quasars. That is, if the $z\sim3$ population
were shifted to $z\sim5$, we should expect a median EW $\approx 70$, but
the median of the observed $z\sim5$ quasars is $\approx 24$. Only 1/10 of
the $M_{1450} \approx -23$ quasars has EW $>40$.

Another check on whether there is evolution in the intrinsic emission 
properties of quasars between $z\sim3$ and $z\sim5$ is to examine the 
profile width of the Ly$\alpha$ emission line. The right panel of 
Figure~\ref{fig:lyafits} shows that the measurements of rHWHM for the two 
redshift ranges are broadly consistent, although this is a noisy measurement. 
We were motivated to perform this check after noticing the narrowness of the 
lines presented by the low-luminosity objects \citep[cf.][]{Matsuoka+17}; 
however, as can be seen in Figure~\ref{fig:lyafits} the $z\sim5$ objects are 
not outliers from the general trend with luminosity observed at $z\sim3$.

\subsection{Implications for completeness}

\begin{figure*}[!ht]
 \epsscale{1.1}
 \plotone{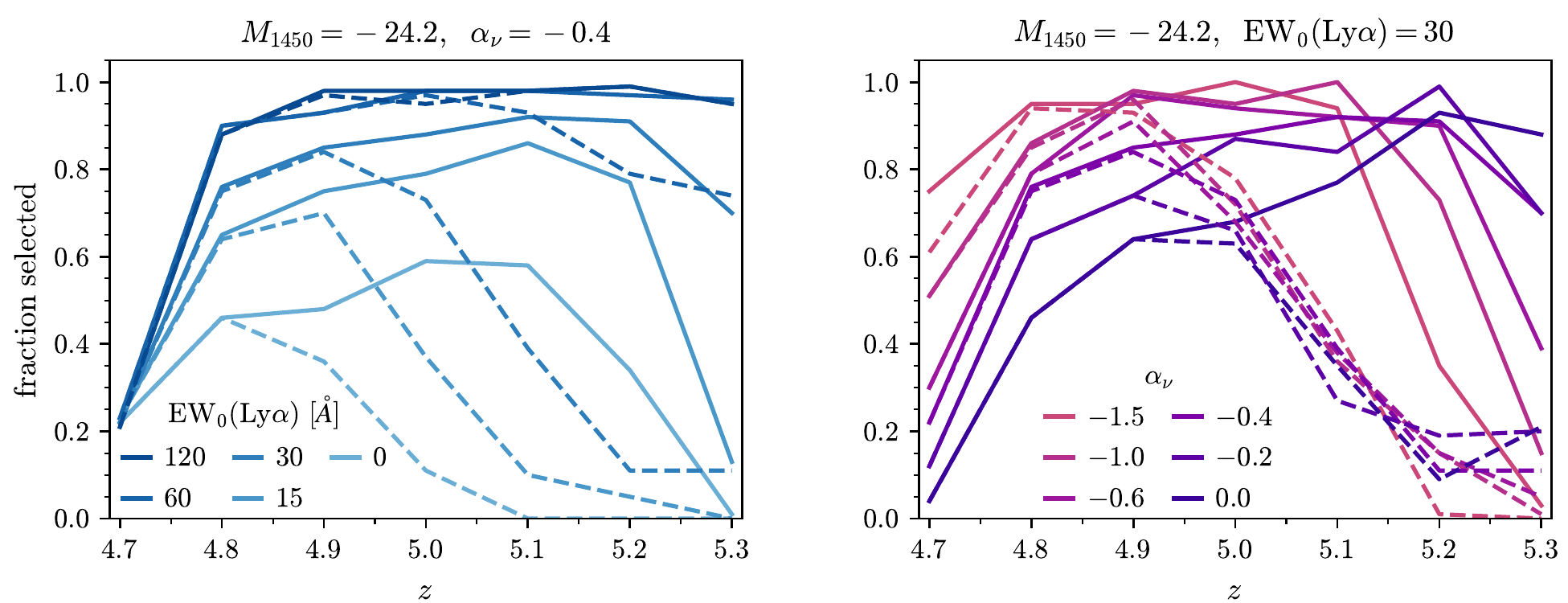}
 \caption{Left: Selection function dependence on Ly$\alpha$ EW for 
 $\alpha_\nu = -0.4$. The line shading represents increasing Ly$\alpha$ EW.
 Solid (dashed) lines are for the weak (strict) color selection criteria.
 At fixed luminosity ($M_{1450}=-24.2$) the selection efficiency for the
 weak cuts drops by a factor of $\sim2$ as the Ly$\alpha$ flux goes to zero. 
 The strict cuts largely exclude $z\ga5$ quasars with modest Ly$\alpha$ EW.
 Right: Selection function at fixed Ly$\alpha$ EW for different UV spectral 
 slopes. In general the dependence on UV slope is much weaker.
 \label{fig:lyasel}
 }
\end{figure*}

We now examine the effect of varying the Ly$\alpha$ emission in our quasar
completness models.  Figure~\ref{fig:lyasel} compares the selection efficiency
along two parameters: Ly$\alpha$ EW and UV continuum slope. We use a grid of 
simulated quasars at $z\sim5$ modeled as a simple power-law continuum with 
index $\alpha_\nu$ and a single Gaussian emission line at 1216\AA. The grid 
includes $\alpha_\nu = (-1.5,-1.0,-0.6,-0.4,-0.2,0.0)$ and
$\EW({\rm Ly}\alpha[\AA]) = (0,15,30,60,120)$. We then run the simulation 
with this simplistic model while allowing for random sampling of the 
Ly$\alpha$ forest transmission and the random scatter of the photometry
\footnote{Details of the simulation, including additional figures, can be 
viewed at 
\url{https://github.com/imcgreer/simqso/blob/master/examples/z5LyaDust.ipynb}}.
The left panel shows the mean selection function at $\alpha_\nu=-0.4$ (a typical 
value for quasars, \citealp{VdB01}) as the Ly$\alpha$ EW is varied. The 
selection function prefers quasars with high EW. Thus it is interesting 
that our observed sample presents a dearth of high EW quasars at $z\sim5$ 
compared to quasars at a similar (low) luminosity at $z\sim3$.

The right panel shows the mean selection function at 
$\EW({\rm Ly}\alpha) = 30$\AA, roughly matching our measurements of 
$z\sim5$ quasars. The dependence on spectral index is less strong, 
although bluer quasars tend to be selected less efficiently at low
redshift and more efficiently at high redshift.

The efficiencies for both the weak (solid lines) and strict (dashed) 
color criteria are presented in Figure~\ref{fig:lyasel}. We have a 
substantial amount of spectroscopic coverage for candidates selected 
by the strict criteria, and for the faint quasar sample the coverage 
is complete. We also select candidates after applying the weak color 
criteria but adding a likelihood cut for both the bright and faint 
candidate samples. Thus our true completeness lies somewhere between 
the dashed and solid curves in Figure~\ref{fig:lyasel}.

Although our selection function clearly depends on the Ly$\alpha$ EW,
a $z=4.9$ quasar with $\EW({\rm Ly}\alpha) = 15$\AA\ has a
$\sim70$\% chance of being selected by either the weak or strict color
criteria. This is the boundary used to define weak-lined quasars by 
\citep{DS09}. A quasar with no Ly$\alpha$ emission has a 
$\sim50$\% ($\sim35$\%) probability of being selected by the weak 
(strict) color criteria.  Unless weak-lined quasars completely dominate 
the population at $z\sim5$, we are likely overestimating our completeness 
at the lowest luminosities by at most a factor of $\sim$2--3. The median 
of our observed sample at $\sim 24$\AA, suggesting that we are not 
overestimating our completeness by such a large factor. Future
surveys less reliant on optical colors will be better suited to address 
this outstanding issue.

\subsection{Additional quasars}

\begin{figure*}[!th]
 \epsscale{1}
 \plotone{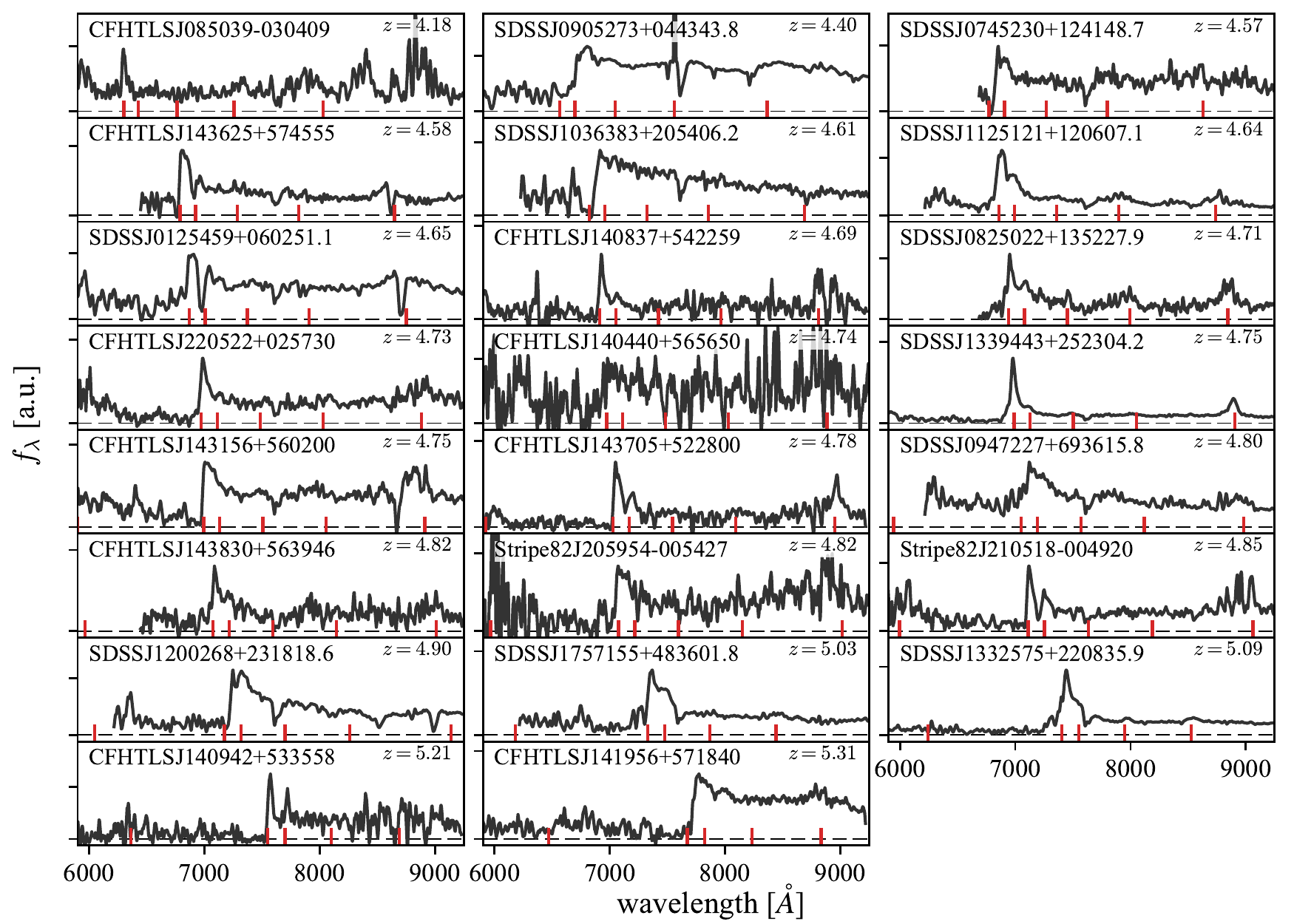}
 \caption{MMT spectra of quasars not included in final sample, corresponding
 entries are in Table~\ref{tab:notcanstab}. Names starting with ``SDSS''
 are from SDSS DR7, ``Stripe82'' from SDSS Stripe 82, and ``CFHTLS'' from
 CFHTLS Wide.
 \label{fig:extraspec}
 }
\end{figure*}

We obtained spectroscopic observations of a number of objects not 
included in our final candidate sample. These can be broken into three 
groups. First, there are 10 $z\sim5$ quasars identified in the CFHTLS 
using early implementations of our color and likelihood selection
methods, but excluded from the final sample. Second, we obtained
spectra for five candidates from our Stripe 82 sample in Paper I
(see Table 3 in that paper) that previously lacked identifications.
Finally, during poor observing conditions we observed bright backup
targets from the SDSS DR7 imaging. The full list of additional quasars 
is given in Table~\ref{tab:notcanstab} and the MMT spectra can be viewed 
in Figure~\ref{fig:extraspec}.

\begin{deluxetable*}{rrcr||rrcr}
 \centering
 \tablecaption{Quasars with spectroscopy not included in final candidate sample\label{tab:notcanstab}}
 \tablewidth{0pt}
 \tablehead{
  \colhead{R.A. (J2000)} & \colhead{Decl. (J2000)} & \colhead{$z$} & \colhead{Source}
  &
  \colhead{R.A. (J2000)} & \colhead{Decl. (J2000)} & \colhead{$z$} & \colhead{Source}
 }
 \startdata
01:25:45.49 & +06:02:51.1 & 4.65 & DR7  &  14:09:42.30 & +53:35:58.2 & 5.21 & CFHT \\
07:45:23.90 & +12:41:48.7 & 4.57 & DR7  &  14:19:56.59 & +57:18:40.4 & 5.31 & CFHT \\
08:25:02.42 & +13:52:27.9 & 4.71 & DR7  &  14:31:56.36 & +56:02:00.8 & 4.75 & CFHT \\
08:50:39.71 & -03:04:09.6 & 4.18 & CFHT  &  14:36:25.45 & +57:45:55.8 & 4.58 & CFHT \\
09:05:27.43 & +04:43:43.8 & 4.40 & DR7  &  14:37:05.17 & +52:28:00.7 & 4.78 & CFHT \\
09:47:22.87 & +69:36:15.8 & 4.80 & DR7  &  14:38:30.83 & +56:39:46.4 & 4.82 & CFHT \\
10:36:38.93 & +20:54:06.2 & 4.61 & DR7  &  17:57:15.95 & +48:36:01.8 & 5.03 & DR7 \\
11:25:12.21 & +12:06:07.1 & 4.64 & DR7  &  20:37:53.64 & +00:05:49.4 &    - & S82 \\
12:00:26.28 & +23:18:18.6 & 4.90 & DR7  &  20:59:54.11 & -00:54:27.7 & 4.82 & S82 \\
13:32:57.45 & +22:08:35.9 & 5.09 & DR7  &  21:00:41.31 & -00:52:03.4 &    - & S82 \\
13:39:44.03 & +25:23:04.2 & 4.75 & DR7  &  21:05:17.97 & -00:49:20.3 & 4.85 & S82 \\
14:04:40.29 & +56:56:50.7 & 4.74 & CFHT  &  21:47:30.25 & +00:12:37.5 &    - & S82 \\
14:08:37.46 & +54:22:59.7 & 4.69 & CFHT  &  22:05:22.15 & +02:57:30.0 & 4.73 & CFHT \\
\enddata

 \tablecomments{The final column lists the imaging source from which the object was selected: DR7 - SDSS DR7, S82 - SDSS Stripe 82, CFHT - CFHTLS Wide. Three non-quasars from the Stripe 82 complete sample given in Paper I are listed with dashes for the redshift entry.}
\end{deluxetable*}

\clearpage

\begin{deluxetable*}{ccrrrrrrrcc}
 \centering
 \tablecaption{Complete sample of quasar candidates with 
$i_{\rm AB}<23.2$ selected from the CFHTLS Wide fields W1, W2, and W3, and 
with partial spectroscopic coverage from Gemini-N, MMT, and LBT.\label{tab:specbright}}
 \tablewidth{0pt}
 \tablehead{
  \colhead{R.A.} & \colhead{Decl.} & \colhead{$g_{\rm AB}$} & \colhead{$r_{\rm AB}$} & \colhead{$i_{\rm AB}$} & \colhead{$z_{\rm AB}$} & \colhead{$\pqso$} & \colhead{$\pqsomidz$} & \colhead{$M_{1450}$} & \colhead{$z$} & \colhead{Tel.}
 }
 \startdata
02:01:00.41 & -05:52:30.2 &   26.04 &   23.30 &   21.67 &   21.72 & 1.0 & 1.0 &         &   &  \\
02:01:02.44 & -05:04:08.6 &   25.59 &   22.94 &   21.17 &   20.88 & 1.0 & 0.00 &         &   &  \\
02:01:53.71 & -05:04:27.2 & $>$26.6 &   24.57 &   22.67 &   22.21 & 1.0 & 0.0 &         &   &  \\
02:01:58.30 & -08:00:48.9 &   25.34 &   23.28 &   21.53 &   21.63 & 1.0 & 1.0 &         &   &  \\
02:02:15.60 & -08:01:21.3 &   27.40 &   25.32 &   22.84 &   22.46 & 1.0 & 0.0 &         &   &  \\
02:03:13.48 & -05:21:37.9 & $>$26.6 &   24.92 &   22.73 &   23.16 & 1.0 & 1.0 &         &   &  \\
02:03:19.34 & -07:55:08.2 &   26.61 &   24.75 &   22.99 &   22.74 & 0.99 & 0.95 &         &   &  \\
02:03:35.87 & -05:03:11.1 & $>$26.6 &   24.15 &   22.82 &   22.62 & 1.0 & 1.0 &         &   &  \\
02:04:03.11 & -06:23:58.2 &   26.09 &   23.73 &   21.98 &   21.83 & 1.0 & 1.0 &         &   &  \\
02:05:33.92 & -07:08:49.1 &   26.61 &   24.78 &   23.07 &   22.59 & 0.00 & 0.0 &         &   &  \\
02:05:41.59 & -03:53:50.8 &   26.45 &   24.64 &   23.16 &   23.10 & 0.97 & 0.06 & -22.9 & 4.60 & Gem \\
02:05:54.37 & -06:26:24.0 &   27.66 &   23.81 &   22.25 &   22.01 & 1.0 & 0.0 &         &   &  \\
02:09:08.47 & -05:01:07.4 &   28.67 &   24.15 &   22.75 &   22.41 & 1.0 & 1.0 &         &   &  \\
02:09:42.92 & -09:27:14.2 &   26.86 &   24.57 &   22.83 &   23.17 & 1.0 & 0.77 &         &   &  \\
02:10:54.68 & -03:42:55.0 & $>$26.1 &   24.64 &   22.68 &   22.23 & 1.0 & 0.0 &         &   &  \\
02:11:01.49 & -03:56:29.0 &   26.91 &   24.77 &   22.96 &   22.46 & 0.18 & 0.00 &         &   &  \\
02:12:47.67 & -05:00:02.8 &   25.20 &   23.16 &   21.79 &   21.70 & 1.0 & 0.0 &         &   &  \\
02:12:57.80 & -08:09:36.8 &   26.35 &   24.37 &   22.65 &   22.20 & 0.0 & 0.0 &         &   &  \\
02:13:05.72 & -10:05:47.9 &   26.43 &   24.43 &   23.04 &   22.81 & 0.99 & 0.01 &         &   &  \\
02:14:46.81 & -09:04:35.2 &   26.48 &   24.60 &   23.09 &   22.85 & 0.71 & 0.00 &         &   &  \\
02:14:48.81 & -08:07:46.7 & $>$26.9 &   25.04 &   23.18 &   22.72 & 0.97 & 0.00 &         &   &  \\
02:15:07.29 & -05:18:49.2 &   26.29 &   24.28 &   22.82 &   22.72 & 1.0 & 0.01 &         &   &  \\
02:15:07.82 & -08:36:59.2 &   25.96 &   23.97 &   22.64 &   22.63 & 1.0 & 0.0 &         &   &  \\
02:15:19.38 & -09:50:18.7 & $>$26.2 &   24.58 &   22.82 &   23.00 & 1.0 & 0.99 &         &   &  \\
02:15:23.27 & -05:29:45.8 &   28.05 &   22.69 &   20.85 &   20.64 & 1.0 & 1.0 & -25.7 & 5.13 & MMT \\
02:16:09.59 & -08:41:41.1 & $>$26.8 &   24.52 &   22.41 &   22.00 & 1.0 & 0.0 &         &   &  \\
02:17:07.49 & -09:54:45.2 &   27.05 &   23.97 &   22.34 &   22.15 & 1.0 & 1.0 &         &   &  \\
02:18:00.49 & -04:47:18.5 &   26.41 &   24.52 &   23.16 &   23.14 & 1.0 & 0.00 &         &   &  \\
02:19:54.18 & -08:29:48.8 & $>$26.8 &   24.57 &   22.45 &   22.21 & 1.0 & 0.99 &         &   &  \\
02:21:12.32 & -03:42:31.6 &   25.82 &   23.59 &   21.42 &   21.48 & 1.0 & 1.0 & -24.9 & 5.02\tablenotemark{a} & MMT \\
02:21:12.61 & -03:42:52.2 &   24.10 &   21.22 &   19.38 &   19.52 & 0.0 & 0.0 & -27.1 & 5.02\tablenotemark{a} & MMT \\
02:23:15.88 & -05:27:30.0 &   26.49 &   23.03 &   21.58 &   21.28 & 1.0 & 1.0 &         &   &  \\
02:24:45.08 & -08:42:28.7 &   25.26 &   22.62 &   20.98 &   20.98 & 1.0 & 1.0 &         &   &  \\
02:25:15.76 & -06:31:15.1 & $>$26.3 &   24.46 &   22.44 &   22.04 & 1.0 & 0.0 &         &   &  \\
02:26:04.73 & -09:18:18.7 &   27.53 &   22.54 &   20.82 &   20.53 & 1.0 & 1.0 &         &   &  \\
02:26:42.39 & -08:38:00.4 &   26.19 &   23.86 &   22.07 &   22.13 & 1.0 & 1.0 &         &   &  \\
02:27:35.87 & -08:08:40.9 &   26.56 &   24.63 &   23.15 &   22.96 & 1.0 & 0.16 & - & - & Gem \\
02:27:40.24 & -07:43:33.8 &   26.64 &   23.32 &   21.73 &   21.34 & 1.0 & 0.0 &         &   &  \\
02:30:59.02 & -06:39:55.7 &   27.47 &   24.46 &   22.99 &   22.74 & 1.0 & 0.32 &         &   &  \\
02:31:19.76 & -08:10:07.1 &   26.03 &   24.17 &   22.67 &   22.42 & 1.0 & 0.0 &         &   &  \\
02:31:37.64 & -07:28:54.4 & $>$26.6 &   21.35 &   19.41 &   19.15 & 0.0 & 0.0 & -27.7 & 5.37 & MMT \\
02:32:45.46 & -06:02:03.1 &   26.32 &   23.47 &   21.87 &   21.80 & 1.0 & 1.0 &         &   &  \\
13:57:47.34 & +53:05:42.5 &   28.27 &   23.32 &   21.29 &   20.85 & 1.0 & 0.0 & -25.6 & 5.32 & MMT \\
13:58:55.97 & +51:43:17.0 &   25.74 &   22.11 &   20.33 &   20.34 & 0.0 & 0.0 & -26.0 & 4.97 & MMT \\
14:01:46.97 & +56:41:44.7 &   25.87 &   23.47 &   21.49 &   21.68 & 1.0 & 0.0 & -24.8 & 4.98 & MMT \\
14:01:49.96 & +51:43:10.3 &   28.28 &   24.54 &   22.85 &   23.05 & 1.0 & 0.99 & -23.0 & 4.20 & MMT \\
14:02:41.71 & +52:39:08.2 &   26.65 &   24.63 &   23.11 &   23.05 & 1.0 & 0.36 & - & - & MMT \\
14:05:17.99 & +54:00:03.2 & $>$26.1 &   25.09 &   23.10 &   22.61 & 1.0 & 0.0 &         &   &  \\
14:06:02.15 & +56:07:19.1 &   28.77 &   24.47 &   23.06 &   22.98 & 1.0 & 0.66 & -23.0 & 4.62 & MMT \\
14:07:10.12 & +51:24:15.1 &   27.00 &   24.56 &   23.08 &   22.63 & 0.42 & 0.31 &         &   &  \\
14:07:29.55 & +53:45:55.7 &   26.46 &   24.15 &   22.77 &   22.37 & 0.36 & 0.36 &         &   &  \\
14:08:22.92 & +53:00:20.9 &   25.11 &   22.75 &   20.99 &   21.16 & 1.0 & 1.0 & -25.5 & 5.07 & MMT \\
14:08:50.27 & +54:41:43.5 &   27.20 &   24.59 &   23.00 &   22.62 & 1.0 & 0.94 &         &   &  \\
14:12:45.95 & +54:46:36.9 &   26.08 &   24.17 &   22.61 &   22.33 & 1.0 & 0.39 &         &   &  \\
14:13:22.22 & +57:45:53.5 &   25.92 &   23.04 &   21.43 &   21.40 & 1.0 & 1.0 & -24.7 & 4.79 & MMT \\
14:13:26.58 & +57:47:06.5 &   27.44 &   23.61 &   21.73 &   21.80 & 1.0 & 1.0 & -24.5 & 4.94 & MMT \\
14:14:04.36 & +51:15:35.8 & $>$26.5 &   23.56 &   21.86 &   21.45 & 1.0 & 0.0 & -24.8 & 5.24 & MMT \\
14:14:31.57 & +57:32:34.1 &   27.61 &   23.37 &   21.73 &   21.51 & 1.0 & 1.0 & -24.8 & 5.16 & MMT \\
14:14:46.82 & +54:46:31.8 &   27.31 &   25.08 &   23.02 &   23.34 & 1.0 & 0.95 & -23.9 & 5.42\tablenotemark{b} & MMT \\
14:18:42.69 & +54:41:31.2 &   26.37 &   23.15 &   21.74 &   21.44 & 1.0 & 1.0 & -24.5 & 4.92 & MMT \\
14:18:45.92 & +54:47:20.7 &   29.06 &   24.50 &   23.18 &   22.95 & 0.99 & 0.62 &         &   &  \\
14:18:58.99 & +54:05:12.2 &   28.08 &   24.75 &   23.11 &   22.63 & 0.73 & 0.16 &         &   &  \\
14:19:29.54 & +51:23:56.1 &   26.74 &   24.56 &   23.05 &   22.74 & 0.34 & 0.21 &         &   &  \\
14:19:56.49 & +55:53:16.2 & $>$26.4 &   25.69 &   23.17 &   22.95 & 1.0 & 0.00 & - & 5.0?\tablenotemark{c} & MMT \\
14:24:08.15 & +51:42:23.8 &   27.03 &   24.59 &   22.78 &   22.29 & 1.0 & 0.0 &         &   &  \\
14:25:19.30 & +51:38:15.6 &   28.35 &   25.04 &   23.15 &   23.29 & 1.0 & 0.98 & -23.0 & 4.84 & MMT \\
14:26:17.28 & +51:55:59.4 &   26.46 &   23.79 &   22.22 &   21.91 & 1.0 & 0.99 & -23.9 & 4.69 & MMT \\
14:26:34.87 & +54:36:22.7 &   23.95 &   21.54 &   19.79 &   20.01 & 0.0 & 0.0 & -26.4 & 4.76 & MMT \\
14:28:53.84 & +56:46:02.1 &   26.07 &   23.76 &   21.99 &   21.84 & 1.0 & 1.0 & -24.1 & 4.73 & MMT \\
14:36:49.39 & +54:15:12.1 &   26.56 &   24.66 &   23.15 &   23.06 & 0.99 & 0.31 &         &   &  \\
14:37:21.78 & +54:28:19.6 &   24.66 &   22.42 &   20.90 &   20.78 & 1.0 & 1.0 &         &   &  \\
14:37:56.54 & +51:51:15.1 & $>$26.6 &   24.25 &   22.37 &   22.21 & 1.0 & 0.98 & -24.2 & 5.17 & MMT \\
14:38:04.05 & +57:36:46.3 &   27.55 &   24.23 &   22.52 &   22.68 & 1.0 & 1.0 & -23.6 & 4.84 & MMT \\
14:38:06.51 & +54:46:03.3 &   25.83 &   23.95 &   22.21 &   21.77 & 1.0 & 0.0 &         &   &  \\
14:39:15.95 & +53:06:22.9 & $>$26.7 &   24.61 &   22.67 &   22.37 & 1.0 & 0.88 & -23.4 & 4.77 & MMT \\
14:39:44.87 & +56:26:26.5 &   27.01 &   24.43 &   22.78 &   22.63 & 1.0 & 0.99 & -23.3 & 4.70 & MMT \\
21:59:55.17 & +01:15:12.9 &   29.20 &   24.64 &   22.96 &   22.70 & 1.0 & 0.38 &         &   &  \\
22:02:33.20 & +01:31:20.3 &   28.16 &   23.95 &   22.09 &   22.02 & 1.0 & 0.14 & -24.6 & 5.23 & LBT \\
22:02:39.31 & +01:03:45.2 &   26.64 &   24.80 &   23.12 &   22.74 & 0.03 & 0.01 &         &   &  \\
22:09:12.11 & +01:32:46.2 &   26.58 &   24.36 &   22.62 &   22.21 & 1.0 & 0.00 &         &   &  \\
22:11:41.01 & +00:11:18.9 & $>$26.7 &   23.99 &   21.91 &   21.73 & 1.0 & 1.0 & -24.8 & 5.23\tablenotemark{d} & MMT \\
22:12:13.43 & -00:04:19.4 &   26.91 &   24.58 &   23.07 &   23.18 & 1.0 & 0.33 &         &   &  \\
22:12:51.49 & -00:42:30.7 &   23.87 &   21.82 &   19.80 &   19.95 & 0.0 & 0.0 & -27.4 & 5.42 & Sloan \\
22:13:05.26 & +00:34:07.8 &   27.45 &   24.46 &   22.76 &   22.26 & 0.92 & 0.0 & -24.0 & 5.31 & MMT \\
22:13:09.67 & -00:24:28.1 &   27.37 &   24.47 &   22.52 &   22.59 & 1.0 & 1.0 & -23.6 & 4.80 & MMT \\
22:14:43.52 & +00:47:33.4 &   25.64 &   23.67 &   22.16 &   22.18 & 1.0 & 1.0 &         &   &  \\
22:15:20.22 & -00:09:08.4 & $>$26.2 &   24.18 &   22.14 &   21.83 & 1.0 & 0.05 & -24.6 & 5.28\tablenotemark{e} & MMT \\
22:16:21.85 & +01:38:14.6 & $>$26.4 &   24.58 &   22.75 &   22.46 & 1.0 & 0.99 & -23.4 & 4.93 & LBT \\
22:16:44.02 & +00:13:48.1 &   25.07 &   22.03 &   20.45 &   20.40 & 1.0 & 0.0 & -25.9 & 5.01 & Sloan \\
22:16:51.92 & +00:28:09.4 &   26.75 &   24.99 &   23.05 &   22.60 & 0.56 & 0.0 &         &   &  \\
22:19:41.90 & +00:12:56.1 &   25.24 &   22.81 &   21.48 &   21.52 & 1.0 & 0.0 & -24.5 & 4.30\tablenotemark{d} & MMT \\
22:19:56.70 & -00:52:40.5 &   27.11 &   24.34 &   22.80 &   22.41 & 0.11 & 0.11 &         &   &  \\
22:19:56.70 & +01:04:29.0 &   26.82 &   24.70 &   23.04 &   22.64 & 0.11 & 0.07 &         &   &  \\
22:20:25.29 & +00:19:30.5 &   26.90 &   24.86 &   23.15 &   22.68 & 0.00 & 0.0 &         &   &  \\
22:20:54.77 & +01:25:53.4 & $>$26.3 &   24.78 &   22.91 &   22.47 & 1.0 & 0.02 & - & - & MMT \\
22:21:13.74 & -00:45:01.2 &   28.89 &   24.65 &   23.02 &   22.52 & 0.58 & 0.32 &         &   &  \\
22:22:16.02 & -00:04:05.6 &   29.58 &   23.69 &   21.88 &   21.84 & 1.0 & 1.0 & -24.4 & 4.95\tablenotemark{d} & MMT
 \enddata
 \tablecomments{Magnitudes are on the AB system \citep{OG83} and have 
been corrected for Galactic extinction. Lower limits on magnitudes are the
50\% completeness limits for the associated CFHTLS pointing from \citet{Gwyn12}. Blank entries indicate lack of spectroscopic coverage and dashes are used for spectroscopic non-quasars.}
 \tablenotetext{a}{Binary quasar, see \citet{McGreer+16}.}
 \tablenotetext{b}{Ly$\alpha$-emitting galaxy, see \citet{McGreer+17}.}
 \tablenotetext{c}{See discussion in \S\ref{sec:mmt}.}
 \tablenotetext{d}{Stripe 82 quasar included in Paper I.}
 \tablenotetext{e}{Also reported in \citet{Ikeda+17}.}

\end{deluxetable*}

\begin{deluxetable*}{ccrrrrrrrc}
 \centering
 \tablecaption{Complete sample of quasar candidates with 
$23.2<i_{\rm AB}<23.7$ selected from the CFHTLS Wide field W1 and with 
partial spectroscopic coverage from Gemini-N. Format is as in Table~\ref{tab:specbright}.\label{tab:specfaint}}
 \tablewidth{0pt}
 \tablehead{
  \colhead{R.A.} & \colhead{Decl.} & \colhead{$g_{\rm AB}$} & \colhead{$r_{\rm AB}$} & \colhead{$i_{\rm AB}$} & \colhead{$z_{\rm AB}$} & \colhead{$\pqso$} & \colhead{$\pqsomidz$} & \colhead{$M_{1450}$} & \colhead{$z$}
 }
 \startdata
02:05:46.63 & -09:55:23.7 & $>$26.6 &   26.18 &   23.55 &   23.37 & 1.0 & 0.10 & -23.3 & 5.37 \\
02:06:18.65 & -04:08:55.5 &   28.46 &   24.94 &   23.32 &   23.10 & 1.0 & 0.87 & -22.8 & 4.87 \\
02:10:56.76 & -05:58:40.4 &   26.87 &   24.93 &   23.42 &   23.11 & 0.05 & 0.00 &         &   \\
02:21:13.96 & -10:09:33.9 &   27.10 &   26.18 &   23.69 &   23.20 & 0.0 & 0.0 & - & - \\
02:26:29.06 & -04:37:04.5 & $>$26.5 &   24.87 &   23.41 &   23.19 & 1.0 & 0.97 & -22.7 & 4.85 \\
02:31:15.46 & -08:19:07.1 &   27.55 &   25.29 &   23.33 &   23.26 & 1.0 & 0.85 & -22.8 & 4.52 \\
02:33:49.79 & -06:45:51.5 & $>$26.5 &   25.25 &   23.52 &   23.05 & 0.54 & 0.01 & - & -
 \enddata

\end{deluxetable*}

\end{document}